\documentclass[10pt,journal, a4paper]{IEEEtran}
\normalsize
\usepackage{graphicx}
\usepackage{makecell}
\usepackage{multicol}
\usepackage{bbm}
\usepackage{lipsum}
\usepackage{diagbox}
\usepackage{blkarray}

\usepackage{hyperref}
\usepackage{float}
\usepackage{url}
\usepackage[utf8]{inputenc} 
\usepackage{array}
\usepackage[utf8]{inputenc} 
\usepackage[T1]{fontenc}
\usepackage{stfloats}
\usepackage{soul}
\usepackage{mathtools}
\usepackage{amsfonts}
\usepackage{tabstackengine}
\usepackage{amssymb}
\usepackage{enumerate}
\usepackage{cite}
\usepackage{calc}
\usepackage{cases}
\usepackage{nicematrix}
\usepackage{color}
\usepackage{epsfig}
\usepackage{setspace}
\usepackage{pstricks}
\usepackage{cancel}
\usepackage{multirow}
\usepackage{mathrsfs}
\usepackage{algorithm}
\usepackage{algpseudocode}
\usepackage{multirow}
\usepackage{romannum}
\usepackage{diagbox}
\algnewcommand{\algorithmicor}{\textbf{ or }}
\algnewcommand{\OR}{\algorithmicor}
\usepackage{booktabs,makecell}
\usepackage{mathrsfs}
\usepackage{enumitem}
\newtheorem{defn}{Definition}
\newtheorem{thm}{Theorem}

\newtheorem{lem}{Lemma}

\newtheorem{remark}{Remark}

\newtheorem{example}{Example}
\graphicspath{{./images/}}
\def\Ddots{\mathinner{\mkern1mu\raise\p@
		\vbox{\kern7\p@\hbox{.}}\mkern2mu
		\raise4\p@\hbox{.}\mkern2mu\raise7\p@\hbox{.}\mkern1mu}}
\makeatother

\begin{document}
	\pagenumbering{arabic}
	\title{Cyclic Wrap-Around Multi-Access Coded Caching with Private Caches}
	\author{\IEEEauthorblockN{Dhruv Pratap Singh, Anjana A. Mahesh and B. Sundar Rajan}\\
		\IEEEauthorblockA{Department of Electrical Communication Engineering, Indian Institute of Science, Bengaluru}\\E-mail: \{dhruvpratap, anjanamahesh, bsrajan\}@iisc.ac.in}
	\maketitle
	\begin{abstract}
	We consider a variant of the coded caching problem where users connect to two types of caches, called private caches and access caches. The problem setting consists of a server having a library of files and a set of access caches. Every user, equipped with a private cache, connects to $L$ neighboring access caches in a cyclic wrap-around fashion. The server populates the private and access caches with file contents in either coded or uncoded format. For this setting, we derive a lower bound on the optimal worst-case transmission rate using cut-set arguments. This lower bound applies to both coded and uncoded placements. We then provide an achievable scheme with uncoded placement and show that our scheme specializes to the well-known Maddah-Ali-Niesen scheme for the dedicated cache network in the absence of access caches. Finally, we show that the proposed scheme achieves optimality in large memory regimes and provide numerical plots comparing the rate of the proposed scheme with the derived lower bound, demonstrating the optimality of our scheme.
\end{abstract}
\section{Introduction}
\textit{Coded Caching}, introduced by Maddah-Ali and Niesen in \cite{MAN}, reduces peak-hour network traffic by utilizing multi-casting opportunities generated as a result of coordinated cache placement at the users. It operates in two phases: the placement phase and the delivery phase. During the placement phase, which occurs when the network load is low, the cache memories in the system are populated with contents either in a coded \cite{CFL}-\cite{AG} or an uncoded fashion \cite{WTP1}, \cite{YMA}, while adhering to their memory constraints. In the delivery phase, when the users reveal their demands, the server seeks to minimize the number of transmissions required to satisfy the demands of all the users. The performance of a coding caching scheme is characterized by the rate-memory trade-off achieved by it, where the rate is the load on the shared link normalized by the file size, and memory refers to the size of the cache memory. The scheme introduced by Maddah-Ali and Niesen \cite{MAN}, referred to as the MAN scheme, addresses the dedicated coded caching problem, where a central server with $N$ files of equal length connects to $K$ users via a shared error-free link. Each user in this network has a cache that can store $M \leq N$ files. The MAN scheme has been proven to achieve the minimum possible rate of transmission \cite{WTP} under uncoded placement, provided $N \geq K$.

The coded caching problem has been studied for various other settings, such as with decentralized placement \cite{MUD}, shared caches \cite{PUE}, \cite{IZY}, secure delivery \cite{STC}, privacy \cite{RPK}, in hierarchical networks \cite{KNAD}, caching where users access multiple caches \cite{RK} - \cite{BE}, caching with offline users \cite{MT}, and many more. Some other works in the coded caching literature that studied settings where users access two different types of caches, one shared between multiple users and the other private to a user, are \cite{MR} - \cite{SMR}. Secretive coded caching, studied in \cite{MR}, used the private caches at the users to store keys required for encryption to ensure secrecy. Coded Caching with shared and private caches, where both the private and shared caches are used to store parts of the files at the server, was studied in \cite{PNR} - \cite{SMR}. 

While the setting considered in this work uses both the access caches and the private caches at the users for storing data files, it differs from the setting in \cite{PNR} and \cite{PNR1} in the number of cache memories accessed by each user. In the systems considered in \cite{PNR, PNR1}, each user accesses one shared cache and its private cache. In contrast, the setting considered in this work involves each user accessing $L\geq 2$ access caches in addition to their private cache.
Furthermore, this work's setting differs from the setting in \cite{SMR} in terms of how the users connect to the access caches and also the number of access caches relative to the number of users. In \cite{SMR}, the users connect to the access caches via a combinatorial topology, where each user connects to a distinct $r$-subset of the $\Lambda$ access caches in the system. Since there is a user connecting to any distinct $r-$subset of the $\Lambda$ access caches, 
there are $K = \binom{\Lambda}{r}$ users in the system. 
In the setting considered in this paper, there are $K$ access caches and $K$ users, and each user connects to $L$ access caches in a cyclic fashion. This topology is referred to as the Cyclic Wrap-around topology. 

This work considers a model where a server with a library of $N$ files connects to $K$ users and $K$ access caches via an error-free wireless link, as shown in Fig \ref{fig1}.. Each user connects to $L$ neighboring access caches in a cyclic wrap-around fashion via error-free wireless links of infinite capacity, as described in \cite{HKD}. However, unlike \cite{HKD}, each user in this model is also equipped with a private cache. This network is a generalization of the multi-access caching network with cyclic wrap-around connectivity \cite{HKD} and the dedicated \cite{MAN} caching network. We refer to this system model as the Cyclic Wrap-Around Multi-Access plus Private (CW-MAP) coded caching setting. To the best of our knowledge, the coded caching problem in this setting has not yet been studied. 

The cyclic wrap-around topology has been extensively studied \cite{RK,HKD,ZWCC,WCWL,NR} in the coded caching literature. Such networks, also called the ring networks, as pointed out in \cite{ZWCC}, are very popular and have been studied in various other contexts as well, like the circular Wyner model for interference networks with limited interference from the neighbors \cite{SSPS,WTS}. The CW-MAP setting is only a natural extension of the cyclic-wraparound network. Further, the assumption that the number of users and the number of access caches in the system are equal may appear to be very limiting. However, when the number of users in the system is larger than the number of caches, we can group the users according to their nearest caches and then, have the number of user groups equal to the number of access caches. Observe that the number of users in each group may vary but all the users within a group will have access to the same similar access cache contents. Then, by keeping the private cache contents also to be the same for all the users in a group, we can repeat the same transmission scheme multiple times such that all the user demands are satisfied, as has been done in \cite{BE}. 

\subsection{Our Contributions}
In this work, we introduce a multi-access coded caching network with cyclic wrap-around connectivity incorporating access caches and private caches at the users. This network can be viewed as a generalization of the dedicated coded caching network presented in \cite{MAN} and the multi-access coded caching network presented in \cite{HKD}. Our contributions are summarized below:
\begin{itemize}
	\item A lower bound on the optimal worst-case rate for the CW-MAP coded caching network, under coded or uncoded placement, is provided using cut-set bound arguments\cite{CT}. Notably, the cut-set bound derived in this work comes out to be the same as the cut-set bound in \cite{SMR}.
	\item A centralized coded caching scheme, under uncoded placement, for the CW-MAP coded caching setting is proposed. The scheme divides each file into subfiles and each subfile further into mini-subfiles and under the placement policy of the proposed scheme, there is no overlap in a users' private cache content and the cache content that user obtains from the access caches it connects to. 
	\item The rate-memory trade-off for the proposed scheme is characterized under the condition that the size of a user's private cache memory is less than the total memory of the access caches that the user connects to.
	\item It is shown that the proposed scheme reverts to the MAN scheme\cite{MAN} when the storage capacity of the access caches is zero.
	\item Optimality of the proposed scheme in the large memory regimes is proved.
	\item Numerical plots are provided to compare the rate attained by the achievability scheme with the lower bound proposed in this paper, demonstrating the optimality of the achievability scheme.
\end{itemize}

\textit{Organization of the paper}:	Section \ref{systemmodel} introduces the system model, outlining the details of the CW-MAP coded caching setting. A brief review of the MAN scheme for dedicated cache networks \cite{MAN} is also provided. The main results of this paper are presented in Section \ref{mainresults}, with detailed proofs provided in Section \ref{proofs} and Section \ref{ssec:optimality}. In Section \ref{numericalcomparison}, we present numerical plots for comparison of the rate achieved by the proposed scheme with the lower bound derived in Section \ref{mainresults}. Finally, we conclude the paper in Section \ref{conclusions}, summarizing our contributions and discussing directions for future work.

\textit{Notation:} We use the notation $\mathbb{Z}^+$ to denote the set of all non-negative integers. The set $\{1,2,\cdots, N\}$ is denoted as $[N]$ for $N \in \mathbb{Z}^+, \ N \geq 1$. The set $\{a,a+1,\cdots,b\}$ where $b\geq a$ is denoted by $[a,b]$, and the set $\{a,a+1,\cdots,b-1\}$ is denoted by $[a,b)$ for $b\geq a+1$, for $a,b\in\mathbb{Z}^+$. The cardinality of a set $A$ is denoted as $|A|$. The notation $<a>_K$ is used to denote the value of $a$ modulo $K$. For an integer $i$, the least integer not less than $i$ is denoted by $\lceil i \rceil$, while $\lfloor i \rfloor$ denotes the largest integer not greater than $i$. For a set $A=\{a_1,a_2,\cdots,a_n\}$, we define $<A>_K=\{<a_1>_K,<a_2>_K,\cdots,<a_n>_K\}$. Given an ordered set $B$ of cardinality $N$, $B(i)$ denotes the $i^{th}$ element of $B$ for $1\leq i\leq N$ and $B(\{i_1,i_2,\cdots, i_n\})=\{B(i_1),B(i_2),\cdots, B(i_n)\}$ for $1\leq i_1,i_2,\cdots,i_n\leq N$. The binomial coefficient $\frac{n!}{k!(n-k)!}$ is denoted as $\binom{n}{k}$ and we assume that $\binom{n}{k}=0$ if $n < 0, k < 0$ or $n<k$. The set of all $t$-sized subsets of a set $S$ of size $|S| \geq t$
, i.e., $\{T: T \subseteq S, |T|=t\}$, is compactly denoted as $\binom{S}{t}$. We use the $\oplus$ symbol to denote the bit-wise XOR operation. 
Finally, the notation $a^+$ is defined as $a^+\triangleq\max(0,a)$.

\section{System Model and Preliminaries}
\label{systemmodel}
We begin this section by introducing the system model. After that, we provide a brief review of the MAN scheme for the dedicated coded caching network \cite{MAN}.
\subsection{System Model}
\begin{figure}
	\centering
	\includegraphics[scale=0.5]{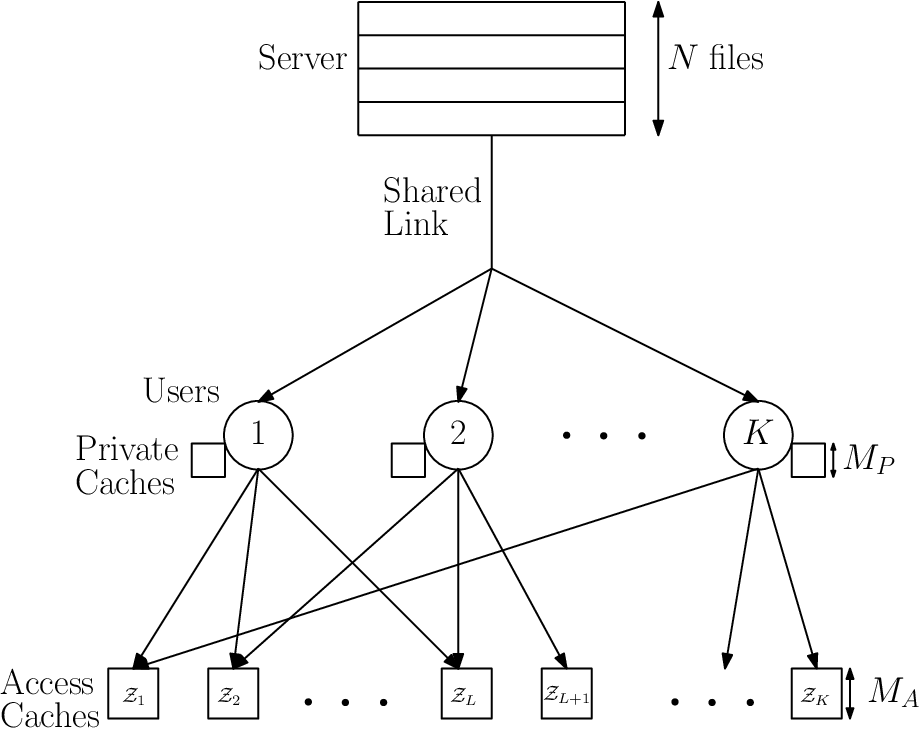}
	\caption{System Model}
	\label{fig1}
\end{figure}
Consider the system model as shown in Fig. \ref{fig1}. A central server has a library of $N$ files, each of size $B$ bits, denoted by $W_1, W_2,\cdots, W_N$. The server connects to $K$ users via an error-free broadcast wireless link, where $N\geq K$. The system consists of $K$ caches capable of storing up to $M_a\leq N$ files. Each of these caches is accessed by multiple users and is referred to as access caches. A user, equipped with a private cache of capacity $M_p\leq N$ files, connects to the access caches in a cyclic wrap-around manner via an infinite capacity, error-free wireless link. Specifically, the $i^{th}$ user, $i\in[ K]$, connects to the set of access caches $<[i,i+L-1]>_K$, where $L$ denotes the access degree. Hence, each user has access to a total memory of $M_aL+M_p$. This model is referred to as the $(K, L, M_a, M_p, N)-$CW-MAP coded caching setting in this paper. The memory pairs $(M_a, M_p)$, which are of interest, satisfy the constraint $0 < M_aL+M_p < N$. 

Notably, when $L=1$, the system reverts to the dedicated coded caching setup with $N$ files and $K$ users, with each user possessing a dedicated cache of capacity $M_a+M_p$. The $(K,L,M_a,M_p,N)-$CW-MAP system operates in two phases:
\begin{figure*}[t]
	\begin{align}
		\label{equationrate}
		&R_{M_a,M_p}=
		\nonumber\\&\frac{(1+\gamma_p)\Big(\binom{K-\gamma_aL}{1+\gamma_p}+(K-1)\binom{K-\gamma_aL-1}{1+\gamma_p}-\sum\limits_{i=1}^{\gamma_aL-1}\binom{K-\gamma_aL-1-(\gamma_aL-i)}{1+\gamma_p-(\gamma_aL-i)}\Big)-\gamma_pK\binom{K-\gamma_aL-2}{1+\gamma_p}-\eta\mathbbm{1}_{\{\gamma_p=\gamma_aL-1\}}}{K\binom{K-\gamma_aL}{\gamma_p}},
	\end{align}where $\eta=(1+\gamma_p)\left(\left(\frac{(K-2\gamma_aL)(K-2\gamma_aL+1)}{2}\right)^++\left((\gamma_aL-1)(K-2\gamma_aL-1)\right)^+\right)-\gamma_pK(K-2\gamma_aL-1)^++\gamma_p\bigg((1+\gamma_aL)(K-2\gamma_aL-1)+\frac{(K-2\gamma_aL-2)(K-2\gamma_aL-1)}{2}\bigg).$\\
	\hrule
\end{figure*}
\begin{enumerate}
	\item \textit{Placement phase}: The server populates the private and access caches with the file contents in either an uncoded fashion, where the parts of the files are copied into the caches without coding across these parts, or in a coded fashion while adhering to their respective memory constraints. The server breaks down the files into subfiles to populate the access caches, and each subfile is further broken down into mini-subfiles to populate the private caches. The caches are populated by the server without the knowledge of the demands of the users. The number of mini-subfiles each file is divided into is called the subpacketization.
	The contents of the access cache $i$ is denoted by $\mathcal{Z}^a_i$, where $i\in[ K]$, while the contents of the private cache of the user ${u}$ is denoted by $\mathcal{Z}^p_{u}$, for $u\in[ K]$. We use the notation $\mathcal{Z}_{u}$ to denote the file contents a user ${u}$ has from its access caches and private cache for $u\in[ K]$. 
	\item \textit{Delivery phase}: Each user $u$ demands one of the $N$ files from the server. Let $W_{d_u}$ denote the file demanded by user $u$, where $d_{u}\in[N]$ and $u\in[K]$. The demands of all the users are represented by the demand vector $\mathbf{d} = (d_{u}: u\in[ K])$. Once the demand vector $\mathbf{d}$ is known to the server, it makes a set of broadcast transmissions ${X}_{\mathbf{d}}$ with a total size of ${R}_{\mathbf{d}}$ files, aiming to satisfy the demands of all the users with a minimum number of transmissions. The rate $R_\mathbf{d}$ is defined as the number of transmissions made by the server for the demand vector $\mathbf{d}$ in the unit of file size. The maximum number of transmissions occur when each user demands a different file; that is, when every element of the demand vector is distinct. This results in the worst-case rate, denoted as $R$.
	\begin{defn}
		For the $(K, L, M_a, M_p, N)-$CW-MAP coded caching setting, a triplet $(M_a, M_p, R)$ is said to be achievable if there exists a coded caching scheme that achieves the worst-case rate $R$ for the memory pair $(M_a, M_p)$, for a large enough file size. We define the optimal worst-case rate for the $(K,L,M_a,M_p,N)-$CW-MAP coded caching setting as:
		\begin{align*}
			R^{\textasteriskcentered}_{M_a,M_p}=\inf\{R:(M_a,M_p,R)\text{ is achievable}\}.
		\end{align*}
	\end{defn}
	The server seeks to design joint placement and delivery policies to achieve $R^{\textasteriskcentered} (M_a, M_p)$.
\end{enumerate}
\subsection{MAN Scheme}
The MAN scheme \cite{MAN} is %defined
 for the dedicated caching network where $K$ users connect to a central server having $N$ files. Every user has a dedicated cache capable of storing $M\leq N$ files. For this setting, a centralized cache placement across the users is made by the server before the file demands of the users are known to it. The delivery phase starts when the server is informed of the demand vector $\mathbf{d}=(d_1,d_2,\cdots,d_K)$, where $d_k$ is the index of the file demanded by the $k^{\text{th}}$ user, $k\in[K]$. The placement and delivery policies in the MAN scheme,and the rate achieved by it are as follows. 
\begin{enumerate}
	\item Placement Phase: Each file is divided into $\binom{K}{t}$ subfiles as $W_n=\{W_{n,\mathcal{T}}:\mathcal{T}\in \binom{[K]}{t} \}$, where $t=\frac{KM}{N}\in \mathbb{Z}^+$. The contents of the cache connected to user $k\in[K]$ is $\mathcal{Z}_k=\{W_{n,\mathcal{T}}:k\in\mathcal{T},\mathcal{T}\in \binom{[K]}{t}, \forall n\in[N]\}$.	
	\item Delivery Phase: The server makes the transmission $X_{\mathcal{S}}$ for every $(t+1)-$subset $\mathcal{S}$ of $[K]$ given by $X_{\mathcal{S}}=\bigoplus\limits_{s\in \mathcal{S}} W_{{d_s},\mathcal{S}\setminus\{s\}}$.
	\item Rate: Each file is divided into $\binom{K}{t}$ subfiles, and a transmission is made for every $(t+1)$-subset of the $K$ users; we have the rate, shown to be optimal under uncoded placement when $N \geq K$, in \cite{YMA}, as $R^{\textasteriskcentered}_D(M)=\frac{\binom{K}{t+1}}{\binom{K}{t}}.$
\end{enumerate}

\section{Main Results}
\label{mainresults}
In this section, we present the main results of this paper. Theorem \ref{thm1} establishes a lower bound on the optimal worst-case rate $R^{\textasteriskcentered}_{M_a,M_p}$ using cut-set arguments \cite{CT}. Theorem \ref{thm2} provides an achievable rate when the placement is uncoded. Finally, Lemma \ref{lem:opt} presents a proof of optimality of the achievability scheme, presented in Theorem \ref{thm2}, for large memory regimes.
\subsection{Lower Bound on the Optimal Rate}
We present a lower bound on the optimal worst-case rate $R^{\textasteriskcentered}_{M_a,M_p}$ for a $(K,L, M_a, M_p, N)-$CW-MAP coded caching system using cut-set arguments as follows:
\begin{thm}
	\label{thm1}
	For the $(K, L, M_a, M_p, N)-$CW-MAP coded caching setting, the following lower bound holds on the optimal worst-case rate:
	\begin{equation}
		R^{\textasteriskcentered}_{M_a,M_p}\geq\max_{s\in[K]} \left(s-\frac{pM_a+sM_p}{\left\lfloor\frac{N}{s}\right\rfloor}\right),
	\end{equation}where $p=\min(s+L-1,K).$
\end{thm}
\begin{IEEEproof}Let $s$ be an integer such that $s\in[K]$. Consider the cache contents $\mathcal{Z}_1,\mathcal{Z}_2,\cdots,\mathcal{Z}_p$, where $p=\min(s+L-1,K)$. For a demand vector $\mathbf{d}_1$, where the first $s$ users demand the first $s$ files and the demands of the other users are arbitrary, the server makes the broadcast transmissions $X_1$. Using $X_1$ and the cache contents $\mathcal{Z}_1,\mathcal{Z}_2,\cdots,\mathcal{Z}_p$, the files $W_1,W_2,\cdots,W_s$ can be decoded. Similarly, for a demand vector $\mathbf{d}_2$, where the first user demands the file $W_{s+1}$, the second user demands the file $W_{s+2}$, and so on, and the $s^{th}$ user demands the file $W_{2s}$, while the demands of the other users are arbitrary, the server makes the broadcast transmissions $X_2$. Using $X_2$ and the cache contents $\mathcal{Z}_1,\mathcal{Z}_2,\cdots,\mathcal{Z}_p$, the files $W_{s+1},W_{s+2},\cdots,W_{2s},$ can be decoded. Continuing in this way, the server makes the transmissions $X_1,X_2,\cdots,X_{\lfloor\frac{N}{s}\rfloor}$. Using these transmissions and the cache contents, the files $W_1, W_2,\cdots, W_{s\lfloor\frac{N}{s}\rfloor},$ files can be decoded. Thus, by the cut-set bound\cite{CT}, we have
	\begin{equation}
		\left\lfloor\frac{N}{s}\right\rfloor R^{\textasteriskcentered}_{M_a,M_p} + sM_p+pM_a\geq s\left\lfloor\frac{N}{s}\right\rfloor
	\end{equation}
	Optimizing over all $s$, we have
	\begin{equation*}
		R^{\textasteriskcentered}_{M_a,M_p}\geq\max_{s\in[K]} \left(s-\frac{pM_a+sM_p}{\left\lfloor\frac{N}{s}\right\rfloor}\right).
	\end{equation*}
\end{IEEEproof}
\begin{remark}
	It is worth noting that the cut-set bound derived above is the same as the one derived in \cite{SMR}. 
\end{remark}
\subsection{Achievable Rate}
The following theorem presents an achievable worst-case rate for the $(K, L, M_a, M_p, N)-$CW-MAP system described in Section \ref{systemmodel}, when $M_aL>M_p$.
\begin{thm}
	\label{thm2} For a $(K,L,M_a,M_p,N)-$CW-MAP coded caching system, a worst-case rate 
	as given in \eqref{equationrate}, is achievable with subpacketization $F=K\binom{K-\gamma_aL}{\gamma_p}$,
	for $\gamma_a=\frac{K M_a}{N}\in[0,K]$ and $\gamma_p=\frac{KM_p}{N}\in[0,\gamma_aL)$.
\end{thm}
\begin{IEEEproof}
	Section \ref{proofs} gives a scheme achieving this rate.
\end{IEEEproof}
Observe that Theorem \ref{thm2} provides rate-memory trade-off for $\gamma_a\in\mathbb{Z}^+$ and $\gamma_p\in\mathbb{Z}^+$. In general, the lower convex envelope of the points in Theorem \ref{thm2} is achievable via memory sharing as explained in subsection \ref{memorysharing}.
\begin{remark}
	It can be observed that while the coded caching scheme proposed in this paper is defined for all values of $M_a$ and $M_p$ in the CW-MAP setting, the scheme in \cite{SMR} is defined for only one value of $M_p$ for the combinatorial MAP setting. Further, while the subpacketization, $F=K\binom{K-\gamma_aL}{\gamma_p}$, for the scheme in this paper is a product of a linearly increasing and an exponentially increasing term in the number of users $K$,  the subpacketization of the scheme in \cite{SMR}, $F=\binom{\Lambda}{t}\binom{\Lambda-t}{r}$, is a product of two exponentially increasing terms in $K$.
\end{remark}
\subsection{Optimality Result}
When the sizes of the access cache memories and the private cache memories satisfy certain constraints, the rate achieved by the coded caching scheme proposed in this paper is shown to be optimal. 
\begin{lem}
	\label{lem:opt}
	For $M_a$ and $M_p$ such that $M_aL+M_p\geq N\left(1-\frac{1}{K}\right)$, the worst-case rate attained by the achievability scheme in section \ref{achievability} is optimal, i.e., $R_{M_a,M_p}=R^{\textasteriskcentered}_{M_a,M_p}$.
	\begin{IEEEproof}
		The proof is deferred till after the achievability scheme is presented and is given in section \ref{ssec:optimality}.
	\end{IEEEproof}
\end{lem}
\section{Achievability Scheme}
\label{proofs}
We begin this section by motivating the main idea behind the achievability scheme using two examples. These examples illustrate all the types of transmissions made by the server. Following this, we present the general placement and delivery scheme that achieves the rate stated in Theorem \ref{thm2}.
\subsection{Motivating Example}
\label{achievability}
\begin{example}
	\label{example1}
	Consider a $(5,2,1,1,5)-$CW-MAP coded caching setting.
	There is a central server, with a library of $N=5$ files, broadcasting to $K=5$ users, each of which are equipped with a private cache of storage capacity of $M_p=1$ file. Each of these users connects to $L=2$ access caches in a cyclic wrap-around manner. User $1$ connects to the access caches $\mathcal{Z}_1^a$ and $\mathcal{Z}_2^a$, user $2$ to access caches $\mathcal{Z}_2^a$ and $\mathcal{Z}_3^a$, and so on till user $5$ which connects to the access caches $\mathcal{Z}_5^a$ and $\mathcal{Z}_1^a$. Each access cache is capable of storing $M_a=1$ file. For this example, $\gamma_a=\frac{KM_a}{N}=\frac{5}{5}=1$. Each file is broken down into $K=5$ subfiles of equal size as $W_n=\{W_{n,\{1\}}, W_{n,\{2\}}, W_{n,\{3\}}, W_{n,\{4\}}, W_{n,\{5\}}\}, \forall n\in[5],$ and placed in the access caches by the central server as listed below:
	\begin{align*}
		&\mathcal{Z}^a_1=\{W_{1,\{1\}},W_{2,\{1\}},W_{3,\{1\}},W_{4,\{1\}},W_{5,\{1\}}\},\\
		&\mathcal{Z}^a_2=\{W_{1,\{2\}},W_{2,\{2\}},W_{3,\{2\}},W_{4,\{2\}},W_{5,\{2\}}\},\\
		&\mathcal{Z}^a_3=\{W_{1,\{3\}},W_{2,\{3\}},W_{3,\{3\}},W_{4,\{3\}},W_{5,\{3\}}\},\\
		&\mathcal{Z}^a_4=\{W_{1,\{4\}},W_{2,\{4\}},W_{3,\{4\}},W_{4,\{4\}},W_{5,\{4\}}\},\\
		&\mathcal{Z}^a_5=\{W_{1,\{5\}},W_{2,\{5\}},W_{3,\{5\}},W_{4,\{5\}},W_{5,\{5\}}\}.
	\end{align*}Each access cache is populated with $5$ subfiles, resulting in each access cache storing $5\times\frac{1}{5}=1$ file, satisfying its memory constraint. Upon connecting to the access caches, users gain access to the subfiles stored in these caches. For example, user $1$ connects to access caches $\mathcal{Z}^a_1$ and $\mathcal{Z}^a_2$ and can access the subfiles $W_{n,\{1\}}$ and $W_{n,\{2\}},$ for $n\in[5]$. However, this style of indexing a subfile does not give any information about the users that have access to a particular subfile. Thus, we replace the subfile-index with a set of cardinality of $\gamma_aL=2$, representing the users that have access to this subfile as shown below:
	\begin{align*}
		&W_{n,1}\rightarrow W_{n,\{1,5\}},\forall n\in[5],\\
		&W_{n,2}\rightarrow W_{n,\{1,2\}},\forall n\in[5],\\
		&W_{n,3}\rightarrow W_{n,\{2,3\}},\forall n\in[5],\\
		&W_{n,4}\rightarrow W_{n,\{3,4\}},\forall n\in[5],\\
		&W_{n,5}\rightarrow W_{n,\{4,5\}},\forall n\in[5].
	\end{align*}Since there are $(K-\gamma_aL)$ users not having access to a particular subfile and we have $\gamma_p=\frac{5}{5}=1$, the server splits each subfile further into $\binom{K-\gamma_aL}{\gamma_p}=\binom{3}{1}=3$ mini-subfiles as \hl{}$W_{n, S}=\{W_{n, S, T}: T\subseteq \binom{[K]\setminus S}{\gamma_p}, |S|=\gamma_aL\}$. We illustrate this splitting below:
	\begin{align*}
		&W_{n,\{1,5\}}=\{W_{n,\{1,5\},\{2\}},W_{n,\{1,5\},\{3\}},W_{n,\{1,5\},\{4\}}\},\\
		&W_{n,\{1,2\}}=\{W_{n,\{1,2\},\{3\}},W_{n,\{1,2\},\{4\}},W_{n,\{1,2\},\{5\}}\},\\
		&W_{n,\{2,3\}}=\{W_{n,\{2,3\},\{1\}},W_{n,\{2,3\},\{4\}},W_{n,\{2,3\},\{5\}}\},\\
		&W_{n,\{3,4\}}=\{W_{n,\{3,4\},\{1\}},W_{n,\{3,4\},\{2\}},W_{n,\{3,4\},\{5\}}\},\\
		&W_{n,\{4,5\}}=\{W_{n,\{4,5\},\{1\}},W_{n,\{4,5\},\{2\}},W_{n,\{4,5\},\{3\}}\},
	\end{align*}$\forall n\in[5]$. The server populates the private caches of the users with parts of subfiles they did not receive upon connecting to the access caches. The contents of the private caches of users is:
	\begin{align*}
		\mathcal{Z}^p_1=&\bigl\{W_{n,\{2,3\},\{1\}},W_{n,\{3,4\},\{1\}},W_{n,\{4,5\},\{1\}},\forall n\in[5]\bigr\},\\
		\mathcal{Z}^p_2=&\bigl\{W_{n,\{3,4\},\{2\}},W_{n,\{4,5\},\{2\}},W_{n,\{1,5\},\{2\}},\forall n\in[5]\bigr\},\\
%	\end{align*}
%	\begin{align*}
		\mathcal{Z}^p_3=&\bigl\{W_{n,\{4,5\},\{3\}},W_{n,\{1,5\},\{3\}},W_{n,\{1,2\},\{3\}},\forall n\in[5]\bigr\},\\
		\mathcal{Z}^p_4=&\bigl\{W_{n,\{1,5\},\{4\}},W_{n,\{1,2\},\{4\}},W_{n,\{2,3\},\{4\}},\forall n\in[5]\bigr\},\\
		\mathcal{Z}^p_5=&\bigl\{W_{n,\{1,2\},\{5\}},W_{n,\{2,3\},\{5\}},W_{n,\{3,4\},\{5\}},\forall n\in[5]\bigr\}.
	\end{align*}Each private cache stores $3$ mini-subfiles of each file, so each private cache stores $\frac{3\times5}{3}=5$ subfiles, which is equivalent to $\frac{5}{5}=1$ file, satisfying its memory constraint. For the convenience of notation, moving forward, the subfile-index set $S$ and the mini-subfile-index set $T$ will be compactly written without the set notation. We now explain how the transmissions are constructed for this placement. 
	
	For the demand vector $\mathbf{d}=(d_{u}:u\in[K])$, consider the mini-subfile $W_{d_2,34,5}$ requested by user $2$. We define the set $I=[2,5]$ by taking the union of the user-index $2$, subfile-index set $34$, and mini-subfile-index set $5$ and rearranging the elements of the set in the increasing order. The position sets $P^I_S$, $P^I_u$, and $P^I_T$ represent the set of positional-indices of the elements in the subfile-index set, the user-index, and the elements in the mini-subfile-index set, respectively, in the set $I$. For $I=[2,5]$, we have $P^I_S=\{2,3\}$, $P^I_u=\{1\}$, and $P^I_T=\{4\}$. 
	
	Now, we check for subsets of $I$ of size $\gamma_aL$ containing consecutive integers other than the subfile index set $S$. Other than $S = 34$, there are two such subsets of $I$ with consecutive integers of cardinality $\gamma_aL=2$% in $I$
	, which are $S_1 = \{2,3\}$ and $S_2 = \{4,5\}$. We increment the elements of $P^I_{S}$ until we obtain the position sets corresponding to the subsets $S_1$ and $S_2$, respectively, with the addition done modulo $|I|$. For instance, $P^I_{S_1} =\{1,2\}$ can be obtained as $P^I_S + 3$, where the addition is done modulo-$4$. Similarly, $P^I_{S_2} =\{3,4\}$ can be obtained as $P^I_S + 1$. Corresponding to each of these subsets $S_1$ and $S_2$, a demanded mini-subfile is obtained by incrementing the user-index-position set $P^I_u$ and mini-subfile-index-position set $P^I_T$ by the same amount. This is demonstrated below:
	\begin{align*}
		&P^I_S=\{2,3\}\xrightarrow{+3}P^I_{S_1}=\{1,2\}\Rightarrow I(\{1,2\})=\{2,3\}\\
		&P^I_u=\{1\}\xrightarrow{+3}P^I_{u_1}=\{4\}\Rightarrow I(\{4\})=\{5\}\\
		&P^I_T=\{4\}\xrightarrow{+3}P^I_{T_1}=\{3\}\Rightarrow I(\{3\})=\{4\}.
	\end{align*}Through this process, we determine the mini-subfile $W_{d_5,23,4}$ for the subfile-index set $23$. To determine the mini-subfile corresponding to the subfile-index set $45$, we follow a similar procedure:
	\begin{align*}
		&P^I_S=\{2,3\}\xrightarrow{+1}P^I_{S_2}=\{3,4\}\Rightarrow I(\{3,4\})=\{4,5\}\\
		&P^I_u=\{1\}\xrightarrow{+1}P^I_{u_2}=\{2\}\Rightarrow I(\{2\})=\{3\}\\
		&P^I_T=\{4\}\xrightarrow{+1}P^I_{T_2}=\{1\}\Rightarrow I(\{1\})=\{2\}.
	\end{align*}Thus, we obtain the mini-subfile $W_{d_3,45,2}$. The server then transmits the coded combination $W_{d_2,34,5}\oplus W_{d_5,23,4}\oplus W_{d_3,45,2}$.					
	Following the above procedure till all the demanded mini-subfiles are exhausted, the server makes the following set of $10$ transmissions:
	\begin{enumerate}
		\item $W_{d_1,23,4}+W_{d_2,34,1}+W_{d_4,12,3}$,
		\item $W_{d_1,23,5}+W_{d_3,51,2}+W_{d_5,12,3}$,
		\item $W_{d_1,34,5}+W_{d_3,45,1}+W_{d_4,51,3}$,
		\item $W_{d_1,34,2}+W_{d_3,12,4}+W_{d_4,23,1}$,
		\item $W_{d_1,45,2}+W_{d_4,12,5}+W_{d_2,51,4}$,
		\item $W_{d_1,45,3}+W_{d_5,34,1}+W_{d_3,51,4}$,
		\item $W_{d_2,51,3}+W_{d_3,12,5}+W_{d_5,23,1}$,
		\item $W_{d_2,34,5}+W_{d_5,23,4}+W_{d_3,45,2}$,
		\item $W_{d_2,45,1}+W_{d_5,12,4}+W_{d_4,51,2}$,
		\item $W_{d_2,45,3}+W_{d_4,23,5}+W_{d_5,34,2}$.
	\end{enumerate}
	It can be verified that all the $K$ users get all the mini-subfiles of their demanded file from the above $10$ transmissions. Since the server makes $10$ transmissions and each file is divided into $15$ mini-subfiles, the normalized rate achieved is $R_{1,1}=\frac{10}{15}=0.667$
\end{example}
In the above Example, whichever mini-subfile we choose, the union of the user-index set, the subfile index set and the mini-subfile index set will contain three subsets of size $\gamma_aL$ containing consecutive integers. Such a scenario, where, upon choosing a demanded mini-subfile $W_{d_u, S, T}$, the union set $I = \{u\} \cup S \cup T$ contains at least one subset of size $\gamma_aL$, other than $S$, containing consecutive integers is termed the \textit{General Case} in the remainder of this paper. 

In general, there will also be mini-subfiles $W_{d_u, S, T}$ such that $S$ is the only subset containing consecutive integers of size $\gamma_aL$ in the union set $I=\{u\}\cup S\cup T$. The transmissions corresponding to such mini-subfiles, which are called \textit{Special Case 1}, are generated differently. 

Further, when $\gamma_p = \gamma_aL-1$, if, for a mini-subfile $W_{d_u, S, T}$, the union set $I = \{u\} \cup S \cup T$ contains only one subset of size $\gamma_aL$ containing consecutive integers, other than $S$, and that subset is $T \cup \{u\}$, such subfiles are also treated differently from the \textit{General Case} and the generation of transmissions corresponding to such subfiles are given as \textit{Special Case 2}. The procedure for generating the transmissions in \textit{Special Case 1} and \textit{Special Case 2} are illustrated with the following example.	
\begin{example}
	\label{example2}
	Consider a $(7,2,1,1,7)-$CW-MAP coded caching system. 			
	The system consists of a central server having a library of $N=7$ files and $K=7$ access caches, each capable of storing $M_a=1$ file. There are $K=7$ users, each equipped with a private cache of size $M_p=1$ file. The users connect to $L=2$ access caches in a cyclic wrap-around fashion. For this system, we have $\gamma_a=\frac{7}{7}=1$. Each file is divided into $K=7$ subfiles as $W_n=\{W_{n,1}, W_{n,2}, W_{n,3}, W_{n,4}, W_{n,5}, W_{n,6}, W_{n,7}\},\forall n\in[7]$, and the placement in the access caches is as described below:
	\begin{align*}
		\mathcal{Z}^a_1=\{W_{n,1},\forall n\in[7]\},\\
		\mathcal{Z}^a_2=\{W_{n,2},\forall n\in[7]\},\\
		\mathcal{Z}^a_3=\{W_{n,3},\forall n\in[7]\},\\
		\mathcal{Z}^a_4=\{W_{n,4},\forall n\in[7]\},\\
		\mathcal{Z}^a_5=\{W_{n,5},\forall n\in[7]\},\\
		\mathcal{Z}^a_6=\{W_{n,6},\forall n\in[7]\},\\
		\mathcal{Z}^a_7=\{W_{n,7},\forall n\in[7]\}.
	\end{align*}Each access cache stores $7$ subfiles, which is equivalent to $\frac{7}{7}=1$ file, satisfying its memory constraint. These subfiles are re-indexed as shown below to indicate which subfiles each user can access:
	\begin{align*}
		W_{n,1}\rightarrow W_{n,71},\forall n\in[7],\\
		W_{n,2}\rightarrow W_{n,12},\forall n\in[7],\\
		W_{n,3}\rightarrow W_{n,23},\forall n\in[7],\\
		W_{n,4}\rightarrow W_{n,34},\forall n\in[7],\\
		W_{n,5}\rightarrow W_{n,45},\forall n\in[7],\\
		W_{n,6}\rightarrow W_{n,56},\forall n\in[7],\\
		W_{n,7}\rightarrow W_{n,67},\forall n\in[7].
			\end{align*}
	Since, $\gamma_p=\frac{7}{7}=1$, each subfile is further divided into $\binom{K-\gamma_aL}{\gamma_p}=\binom{7-2}{1}=5$ mini-subfiles as shown below:
	\begin{align*}
		W_{n,12}=&\bigl\{W_{n,12,3},W_{n,12,4},W_{n,12,5},W_{n,12,6},W_{n,12,7}\bigr\},\\
		W_{n,23}=&\bigl\{W_{n,23,1},W_{n,23,4},W_{n,23,5},W_{n,23,6},W_{n,23,7}\bigr\},\\
		W_{n,34}=&\bigl\{W_{n,34,1},W_{n,34,2},W_{n,34,5},W_{n,34,6},W_{n,34,7}\bigr\},\\
		W_{n,45}=&\bigl\{W_{n,45,1},W_{n,45,2},W_{n,45,3},W_{n,45,6},W_{n,45,7}\bigr\},\\
		W_{n,56}=&\bigl\{W_{n,56,1},W_{n,56,2},W_{n,56,3},W_{n,56,4},W_{n,56,7}\bigr\},\\
		W_{n,67}=&\bigl\{W_{n,67,1},W_{n,67,2},W_{n,67,3},W_{n,67,4},W_{n,67,5}\bigr\},\\
		W_{n,71}=&\bigl\{W_{n,71,2},W_{n,71,3},W_{n,71,4},W_{n,71,5},W_{n,71,6}\bigr\},
	\end{align*}$\forall n\in[7]$. The placement made in the private caches of the users is as shown below:
	\begin{align*}
		\mathcal{Z}^p_{1}=&\bigl\{W_{n,23,1},W_{n,34,1},W_{n,45,1},W_{n,56,1},W_{n,67,1}\bigr\},\\
		\mathcal{Z}^p_{2}=&\bigl\{W_{n,34,2},W_{n,45,2},W_{n,56,2},W_{n,67,2},W_{n,71,2}\bigr\},\\
		\mathcal{Z}^p_{3}=&\bigl\{W_{n,12,3},W_{n,45,3},W_{n,56,3},W_{n,67,3},W_{n,71,3}\bigr\},\\
		\mathcal{Z}^p_{4}=&\bigl\{W_{n,12,4},W_{n,23,4},W_{n,56,4},W_{n,67,4},W_{n,71,4}\bigr\},\\
		\mathcal{Z}^p_{5}=&\bigl\{W_{n,12,5},W_{n,23,5},W_{n,34,5},W_{n,67,5},W_{n,71,5}\bigr\},\\
		\mathcal{Z}^p_{6}=&\bigl\{W_{n,12,6},W_{n,23,6},W_{n,34,6},W_{n,45,6},W_{n,71,6}\bigr\},\\
		\mathcal{Z}^p_{7}=&\bigl\{W_{n,12,7},W_{n,23,7},W_{n,34,7},W_{n,45,7},W_{n,56,7}\bigr\},
	\end{align*}
$\forall n\in[7]$. Each private cache stores $35$ mini-subfiles, equivalent to $\frac{35}{35}=1$ file, satisfying its memory constraint. Since Example \ref{example1} demonstrated the construction of transmissions for the \textit{General Case}, here, we only list all the transmissions made by the server for the \textit{General Case} and do not explain how these transmissions are obtained. Subsequently, we explain how transmissions are constructed for \textit{Special Case 1} and \textit{Special Case 2}. For the demand vector $\mathbf{d}=(d_1,d_2,\cdots,d_7)$, the transmissions made by the server for the \textit{General Case} are as follows:
	\begin{enumerate}
		\item $W_{d_{1},23,4}\oplus W_{d_{4},12,3}\oplus W_{d_{2},34,1}$
		\item $W_{d_{1},23,5}\oplus W_{d_{5},12,3}$
		\item $W_{d_{1},23,6}\oplus W_{d_{6},12,3}$
		\item $W_{d_{1},23,7}\oplus W_{d_{7},12,3}\oplus W_{d_{3},71,2}$
		\item $W_{d_{1},34,2}\oplus W_{d_{3},12,4}\oplus W_{d_{4},23,1}$
		\item $W_{d_{1},34,5}\oplus W_{d_{3},45,1}$
		\item $W_{d_{1},45,3}\oplus W_{d_{5},34,1}$
		\item $W_{d_{1},45,6}\oplus W_{d_{4},56,1}$
		\item $W_{d_{1},56,4}\oplus W_{d_{6},45,1}$
		\item $W_{d_{1},56,7}\oplus W_{d_{5},67,1}\oplus W_{d_{6},71,5}$
		\item $W_{d_{1},67,2}\oplus W_{d_{6},12,7}\oplus W_{d_{2},71,6}$
		\item $W_{d_{1},67,3}\oplus W_{d_{3},71,6}$
		\item $W_{d_{1},67,4}\oplus W_{d_{4},71,6}$
		\item $W_{d_{1},67,5}\oplus W_{d_{7},56,1}\oplus W_{d_{5},71,6}$
		\item $W_{d_{2},34,5}\oplus W_{d_{5},23,4}\oplus W_{d_{3},45,2}$
		\item $W_{d_{2},34,6}\oplus W_{d_{6},34,2}$
		\item $W_{d_{2},34,7}\oplus W_{d_{7},23,1}$
		\item $W_{d_{2},45,3}\oplus W_{d_{4},23,5}\oplus W_{d_{5},34,2}$
		\item $W_{d_{2},45,6}\oplus W_{d_{4},56,2}$
		\item $W_{d_{2},56,4}\oplus W_{d_{6},45,2}$
		\item $W_{d_{2},56,7}\oplus W_{d_{5},67,2}$
		\item $W_{d_{2},67,1}\oplus W_{d_{6},71,2}\oplus W_{d_{7},12,6}$
		\item $W_{d_{2},67,5}\oplus W_{d_{7},56,2}$
		\item $W_{d_{2},71,3}\oplus W_{d_{3},12,7}\oplus W_{d_{7},23,1}$
		\item $W_{d_{2},71,4}\oplus W_{d_{4},12,7}$
		\item $W_{d_{2},71,5}\oplus W_{d_{5},12,7}$
		\item $W_{d_{3},12,5}\oplus W_{d_{5},23,1}$
		\item $W_{d_{3},12,6}\oplus W_{d_{6},23,1}$
		\item $W_{d_{3},45,6}\oplus W_{d_{6},34,5}\oplus W_{d_{4},56,3}$
		\item $W_{d_{3},45,7}\oplus W_{d_{7},34,5}$
		\item $W_{d_{3},56,4}\oplus W_{d_{5},34,6}\oplus W_{d_{6},45,3}$
		\item $W_{d_{3},56,7}\oplus W_{d_{5},67,3}$
		\item $W_{d_{3},67,1}\oplus W_{d_{6},71,3}$
		\item $W_{d_{3},67,5}\oplus W_{d_{7},56,3}$
		\item $W_{d_{4},23,6}\oplus W_{d_{6},34,2}$
		\item $W_{d_{4},23,7}\oplus W_{d_{7},34,2}$
		\item $W_{d_{4},56,7}\oplus W_{d_{7},45,7}\oplus W_{d_{5},67,4}$
		\item $W_{d_{4},67,1}\oplus W_{d_{6},71,4}$
		\item $W_{d_{4},67,5}\oplus W_{d_{6},45,7}\oplus W_{d_{7},56,4}$
		\item $W_{d_{4},71,2}\oplus W_{d_{7},12,4}$
		\item $W_{d_{5},34,7}\oplus W_{d_{7},45,3}$
		\item $W_{d_{5},71,2}\oplus W_{d_{7},12,5}$
	\end{enumerate}
	We now introduce the \textit{Special Case 1}. This case deals with mini-subfiles $W_{d_{u}, S, T}$ where the set $I = \{u\} \cup S \cup T$ consists of only one subfile-index set, namely $S$. For instance, consider the mini-subfile $W_{d_{1},34,6}$. Here, $I = \{1, 3, 4, 6\}$, with $S = 34$. The server replaces the user index $1$ with an element from the mini-subfile-index set $6$, resulting in $W_{d_{6},34,1}$. Then, it performs a bit-wise XOR operation on these mini-subfiles to generate the transmission $W_{d_{1},34,6} \oplus W_{d_{6},34,1}$. All transmissions made by the server for \textit{Special Case 1} are listed below:
	\begin{enumerate}
		\item $W_{d_{1},34,6}\oplus W_{d_{6},34,1}$
		\item $W_{d_{1},56,3}\oplus W_{d_{3},56,1}$
		\item $W_{d_{2},45,7}\oplus W_{d_{7},45,2}$
		\item $W_{d_{2},67,4}\oplus W_{d_{4},67,2}$
		\item $W_{d_{3},71,5}\oplus W_{d_{5},71,3}$
		\item $W_{d_{4},12,6}\oplus W_{d_{6},12,4}$
		\item $W_{d_{5},23,7}\oplus W_{d_{7},23,5}$
	\end{enumerate}
	We now explain \textit{Special Case 2}, which corresponds to the mini-subfiles $W_{d_{u}, S, T}$ such that the set $I=\{u\}\cup S\cup T$ has only two subfile index sets, which are $S$ and $\{u\}\cup T$. Note that such a case arises only when $\gamma_p = \gamma_aL-1$. Consider the mini-subfile $W_{d_{1},34,7}$. For this mini-subfile, we have $I=\{1,3,4,7\}$, which has two subfile-index sets, $34$ and $71$. Hence, following the procedure in \textit{General Case}, the server performs the bit-wise XOR operation of mini-subfiles $W_{d_{1},34,7}$ and $W_{d_{4},71,3}$. However, in \textit{Special Case 2}, in addition to the two sub-files generated by the procedure in the \textit{General Case}, the server also replaces the user-index with each element of the mini-subfile-index set for both the mini-subfiles $W_{d_{1},34,7}$ and $W_{d_{4},71,3}$ to obtain the mini-subfiles $W_{d_{7},34,1}$ and $W_{d_{3},71,4}$ respectively. Finally, the server performs the bit-wise XOR operation of all these mini-subfiles to obtain the transmission $W_{d_{1},34,7}\oplus W_{d_{4},71,3}\oplus W_{d_{7},34,1}\oplus W_{d_{3},71,4}$. We provide all the transmissions made by the server for \textit{Special Case 2} below:
	\begin{enumerate}
		\item $W_{d_{1},34,7}\oplus W_{d_{4},71,3}\oplus W_{d_{3},71,4}\oplus W_{d_{7},34,1}$
		\item $W_{d_{1},45,2}\oplus W_{d_{4},12,5}\oplus W_{d_{5},12,4}\oplus W_{d_{2},45,1}$
		\item $W_{d_{1},45,7}\oplus W_{d_{5},71,4}\oplus W_{d_{4},71,5}\oplus W_{d_{7},45,1}$
		\item $W_{d_{1},56,2}\oplus W_{d_{5},12,6}\oplus W_{d_{6},12,5}\oplus W_{d_{2},56,1}$
		\item $W_{d_{2},56,3}\oplus W_{d_{5},23,6}\oplus W_{d_{6},23,5}\oplus W_{d_{3},56,2}$
		\item $W_{d_{2},67,3}\oplus W_{d_{7},23,6}\oplus W_{d_{6},23,7}\oplus W_{d_{3},67,2}$
		\item $W_{d_{3},67,4}\oplus W_{d_{7},34,6}\oplus W_{d_{6},34,7}\oplus W_{d_{4},67,3}$
	\end{enumerate}
	It can be verified that all the demanded mini-subfiles are covered by a transmission in either the \textit{General Case}, \textit{Special Case 1} or \textit{Special Case 2} and that every user can recover all the mini-subfiles of its requested file. Since the server makes a total of $42+7+7=56$ transmissions to satisfy the demands of all the users and the subpacketization is $F=35$, the rate achieved by the coded caching scheme is $R_{1,1}=\frac{56}{35}=1.6$.
\end{example}
We now present the general placement and delivery policy for the achievability scheme.
\subsection{Achievability Scheme}
\textit{Placement Phase}: First, we describe the placement policy of the access caches. Each file is split into $K$ non-overlapping subfiles of equal size as follows:
\begin{equation}
	\label{subfileconstruction}
	W_n=\{W_{n,i}:i\in[K]\},\;\forall n\in[N].
\end{equation}
The contents of the access cache $k\in[K]$ are given as:
\begin{equation}
	\label{placementaccesscaches}\mathcal{Z}^a_k=\{W_{n,<k+(j-1)L>_K}:j\in[\gamma_a],\forall n\in[N]\},
\end{equation}where $L$ is the access degree and, $\gamma_a=\frac{KM_a}{N}\in\mathbb{Z}^+$, is the access cache memory replication factor. This is the same placement strategy as described in \cite{WCWL}. Each access cache is populated by $N\gamma_a=\frac{N\frac{KM_a}{N}}{K}=M_a$ files, satisfying its memory constraint. % of the access caches. 
However, as explained previously, this indexing policy of the subfiles does not intuitively specify which users will have access to which subfiles. In order to rectify this, we re-index the subfile $W_{n,i}$ as $W_{n,S=<[i-\gamma_aL+1,i]>_K}$, where the subfile-index set $S=<[i-\gamma_aL+1,i]>_K$ of cardinality $\gamma_aL$ describes the users which have access to the subfile $W_{n,S}$. Moving forward, we refer to any set of the form $<[j-\gamma_aL+1,j]>_K$, for some $j\in[K]$, as a subfile-index set.

Now, we describe the placement strategy of the private caches. For a user $u\in[K],$ the server populates its private cache with mini-subfiles of the subfiles that it does not obtain from the access caches it connects to. Each subfile $W_{n,S}$ is wanted by $K-\gamma_aL$ users and hence, is further divided into $\binom{K-\gamma_aL}{\gamma_p}$ mini-subfiles as:
\begin{align}
	\label{minisubfileconstruction}
	W_{n,S}=\bigg\{W_{n,S,T}:T\in\binom{[K]\setminus S}{\gamma_p}\bigg\},
\end{align} where, $\gamma_p=\frac{KM_p}{N}\in\mathbb{Z}^+$, is the private cache memory replication factor. The mini-subfile of the subfile ${S}$ of the file $n$, stored in the private cache of $u$, is denoted as $\{W_{n,S,T}:u\in T,u\not\in S\}$. The contents of the private cache of user $u$ are given as:
\begin{equation}
	\label{placementpolicyprivatecaches}
	\begin{split}
		\mathcal{Z}^p_u=&\{W_{n,S,T}:|S|=\gamma_aL, u\not\in S,T\in\binom{[K]\setminus S}{\gamma_p}, \\ &u\in T,\forall n\in[N]\},
	\end{split}
\end{equation}where $S$ is a subfile-index set.
Each private cache stores $\frac{N(K-\gamma_aL)\binom{K-\gamma_aL-1}{\gamma_p-1}}{K\binom{K-\gamma_aL}{\gamma_p}}=\frac{N\gamma_p}{K}=M_p$ files, satisfying its memory constraint. %of the private caches. 
Under the outlined placement policies, there is no overlap in the contents of a user's private cache and the access caches it connects to.

\begin{algorithm*}[t]
	\caption{Algorithm for generating transmission during delivery phase}
	\label{Algo1}
	\hspace*{\algorithmicindent} \textbf{Input:} $\mathbf{d}=(d_{u}:u\in[K])$, $\mathcal{Z}=(\mathcal{Z}_{u}:u\in[K])$. \\
	\hspace*{\algorithmicindent} \textbf{Output:} The set of transmissions $X$.
	\begin{algorithmic}[1]
		\State \label{line1}Initialize $X=\emptyset$.
		\State \label{line2}For each user $u$, define the user-demand set $D_u=\{(S,T):S\text{ is a subfile-index set},u\not\in S,T\in\binom{[K]\setminus \{\{u\}\cup S\}}{\gamma_p}\}$.
		\For{$u\in[K]$}\label{line3}
		\While{${D}_u\not=\emptyset$}\label{line4}
		\State \label{line5}Select an element $(S,T)$ from ${D}_u$.
		\State \label{line6}For $(S,T)$, define ${I}=\{u\}\cup S\cup T$, such that the elements of $I$ are arranged in an increasing order.
		\If{$I$ contains only one subfile-index set, which is $S$,}\Comment{Special Case 1}\label{line7}
		\State \label{specialcase1}$X^\prime=W_{d_{u},S,T}\bigoplus\limits_{t\in T} W_{d_{t},S,\{\{u\}\cup T\}\setminus\{t\}}$.
		\ElsIf{$I$ contains two subfile-index sets, which are $S$, and, $\{u\}\cup T$,}\Comment{Special Case 2}\label{line8}
		\State \label{line9}Let $j\in[2\gamma_aL-1]$ be such that $I\left(P^I_S+j\right)=\{u\}\cup T$. 
		\State \label{specialcase2}\begin{align*}X^\prime=W_{d_{u},S,T}\bigoplus\limits_{t\in T} W_{d_{t},S,\{\{u\}\cup T\}\setminus\{t\}}\oplus W_{d_{I(P^I_u+j)},I(P^I_S+j),I(P^I_T+j)}\bigoplus\limits_{i\in I(P^I_T+j)} W_{d_{i},I(P^I_S+j),\{I(P^I_T+j)\cup I(P^I_u+j)\}\setminus \{i\}}.\end{align*}
		\Else\Comment{General Case}
		\State \label{generalcase}$X^\prime=W_{d_{u},S,T}\bigoplus\limits_{i\in[\gamma_aL+\gamma_p]} \{W_{d_{I(P^I_u+i)},I(P^I_S+i),I(P^I_T+i)}:I(P^I_S+i)\text{ is a subfile-index set}\}$
		\EndIf
		\State \label{line10}$X\gets X\cup X^\prime$.
		\State \label{line11}Let $S_c=\{W_{d_{u},{S},{T}}:W_{d_{u},{S},{T}}\text{ is a mini-subfile in $X^\prime$}\}$. For each $W_{d_{\hat{u}},\hat{{S}},\hat{{T}}}\in S_c$, do ${D}_{\hat{u}}\gets{D}_{\hat{u}}\setminus~(\hat{{S}},\hat{{T}})$
		\EndWhile
		\EndFor
	\end{algorithmic}
	\noindent\rule{\textwidth}{1pt}
\end{algorithm*}
Before explaining the delivery phase, we formally define the positional index notations, $P^I_S$, referred to as the subfile-index-position set in $I$, $P^I_u$, referred to as the user-index-position in $I$, and $P^I_T$, referred to as the mini-subfile-index-position set in $I$. After defining $P^I_u$, $P^I_S$, and $P^I_T$, we give an illustrative example explaining these notations.
\begin{defn}
	For a mini-subfile $W_{d_{u}, S, T}$, define the set $I\triangleq\{u\}\cup S\cup T$, such that the elements of $I$ are arranged in an increasing order. For the set $I$, the user-index-position set $P^I_u$, the subfile-index-position set $P^I_S$ and the mini-subfile-index-position set $P^I_T$, respectively, are defined as:
	\begin{align*}
		P^I_u\triangleq\{i:I(i)=u\},\\
		P^I_S\triangleq\{i:I(i)\in S\},\\
		P^I_T\triangleq\{i:I(i)\in T\}.
	\end{align*}In other words, $P^I_u$ is the singleton set containing the positional index of the element $u$ in the set $I$, and $P^I_S$ and $P^I_T$ are the sets containing the positional indices of the elements in $S$ and $T$, respectively, in the set $I$ when the elements of $I$ are arranged in an increasing order. Further, addition of an integer $j$ to $P^I_u$, $P^I_S$, and $P^I_T$, is defined as:
	\begin{align*}
		P^I_u+j\triangleq\{<p+j>_{|I|}: p\in P^I_u\},\\
		P^I_S+j\triangleq\{<p+j>_{|I|}:p\in P^I_S\},\\
		P^I_T+j\triangleq\{<p+j>_{|I|}:p\in P^I_T\}.
	\end{align*} 
\end{defn}
\begin{example}
	Consider a mini-subfile $W_{d_{3},56,2}$. For this mini-subfile, we have $I=\{2,3,5,6\}$. For this $I$, we have $P^I_u=\{2\}, P^I_S=\{3,4\},\text{ and, }P^I_T=\{1\}$. Adding two to $P^I_u$, $P^I_S$, and $P^I_T$, we have $P^I_u+2=\{<2+2>_4\}=\{4\}$, $P^I_S+2=\{<3+2>_4,<4+2>_4\}=\{1,2\}$, and $P^I_T+2=\{<1+2>_4\}=\{3\}$. Thus, we have $I(P^I_u+2)=I(\{4\})=\{6\}$, $I(P^I_S+2)=I(\{1,2\})=\{2,3\}$, and $I(P^I_T+2)=I(\{3\})=\{5\}$.
\end{example}
\textit{Delivery Phase}: For the demand vector $\mathbf{d}$, the server broadcasts the set of transmissions $X$ returned by Algorithm \ref{Algo1}. The working of Algorithm \ref{Algo1} is explained below.

Algorithm \ref{Algo1} starts by initializing the set of transmissions $X$ to an empty set. Then, for each user $u$, its user-demand set, defined as the set of indices of the mini-subfiles wanted by it, is constructed. For instance, in Example \ref{example2},  the user-demand set corresponding to user $1$ is  $D_1=\bigl\{(23,4),(23,5),(23,6),(23,7),(34,2),(34,5),\\(34,6),(34,7),(45,2),(45,3),(45,6),(45,7),(56,2),\\(56,3),(56,4),(56,7),(67,2),(67,3),(67,4),(67,5)\bigr\}$. Algorithm \ref{Algo1} starts with the first user and picks an element $(S, T)$ from its user-demand set. For this element, Algorithm \ref{Algo1} calculates the set $I = \{u\} \cup S \cup T$ (Line \ref{line6}). Depending on the subfile-index sets contained in the set $I$, Algorithm \ref{Algo1} constructs a transmission involving the chosen mini-subfile as described in either Line \ref{specialcase1}, Line \ref{specialcase2} or Line \ref{generalcase}. 

Line \ref{specialcase1} describes the construction of a transmission corresponding to \textit{Special Case 1} where the set $I$ has only one subfile-index set, which is $S$. Line \ref{specialcase2} describes the transmission for \textit{Special Case 2} where the set $I$ has two subfile-index sets, namely, $S$ and $\{u\}\cup T$. Note that \textit{Special Case 2} arises only when $\gamma_p = \gamma_aL-1$. Finally, Line \ref{generalcase} describes the transmissions when none of the above conditions are applicable, which is referred to as the \textit{General Case}. Let us say, in Example \ref{example2}, Algorithm \ref{Algo1} picks the user $1$ and picks the entry $(34,6)$ from the user-demand set $D_1$ of user $1$. For this entry, the set $I=\{1,3,4,6\}$ has only one subfile-index set, $S=34$. Thus, Algorithm \ref{Algo1} constructs the transmission $W_{d_{1},34,6}\oplus W_{d_{6},34,1}$ as described in Line \ref{specialcase1}. Meanwhile, when Algorithm \ref{Algo1} selects the element $(34,7)$ from the user-demand set $D_1$, the set $I=\{1,3,4,7\}$ has two subfile-index sets, $S=34$ and $\{u\}\cup T=71$ which corresponds to \textit{Special Case 2}. Therefore, Algorithm \ref{Algo1} constructs the transmission $W_{d_{1},34,7}\oplus W_{d_{7},34,1}\oplus W_{d_{3},71,4}\oplus W_{d_{4},71,3}$ in accordance with Line \ref{specialcase2}. Finally, when Algorithm \ref{Algo1} picks the entry $(23,4)$ from the user-demand set of user $1$, the conditions mentioned in Line \ref{line7} and Line \ref{line8} do not apply. Thus, Algorithm \ref{Algo1} constructs the transmission $W_{d_{1},23,4}\oplus W_{d_{4},12,3}\oplus W_{d_{2},34,1}$ as described in Line \ref{generalcase}.

As each transmission is constructed, it is added to the set of transmissions $X$ that will be output by the Algorithm, and the indices of the mini-subfiles occurring in the transmission are removed from the user-demand set of the user who requested those mini-subfiles (Line \ref{line11}). This process is repeated for all the elements in the user-demand sets of all the users. 

\textit{Decodability}: A mini-subfile $W_{d_{u}, S, T}$ can be included in any one of the three types of transmissions generated by Algorithm \ref{Algo1}, which is determined by its corresponding union set $I=\{u\}\cup S\cup T$. We designate the transmission specified in Line \ref{specialcase1} as \textit{Special Case 1}, that in Line \ref{specialcase2} as \textit{Special Case 2}, and the one in Line \ref{generalcase} as \textit{General Case}. The decodability of each case is discussed below. 
\begin{enumerate}
	\item \textit{Special Case 1}: Consider the mini-subfile $W_{d_{u}, S, T}$ demanded by the user $u$ in a transmission $X^\prime$ given by Line \ref{specialcase1} of Algorithm \ref{Algo1}. In this transmission, every mini-subfile, other than $W_{d_{u}, S, T}$, possesses a mini-subfile-index set that includes user $u$, i.e., for every mini-subfile $W_{d_{\hat{u}},\hat{S},\hat{T}}\neq W_{d_{u},S,T}$ in $X^\prime$, we have $u\in\hat{T}$. As a result, user $u$ has access to all the mini-subfiles within the transmission, except $W_{d_{u}, S, T}$. Thus, user $u$ can decode its requested mini-subfile $W_{d_{u}, S, T}$ from the transmission $X^\prime$.
	\item \textit{Special Case 2}: Consider a transmission $X^\prime$, generated according to Line \ref{specialcase2} of Algorithm \ref{Algo1}, that includes a mini-subfile $W_{d_{u},S,T}$ which is demanded by user $u$. In $X^\prime$, every mini-subfile of the form $W_{d_{t}, S,{\{u\}\cup T}\setminus\{t\}}$ is available to user $u$, since $u\in \{\{u\}\cup T\}\setminus \{t\}$. Additionally, the mini-subfile $W_{d_{I(P^I_u+j)},I(P^I_S+j),I(P^I_T+j)}$ and all the mini-subfiles included in $\bigoplus\limits_{i\in I(P^I_T+j)} W_{d_{i},I(P^I_S+j),{I(P^I_T+j)\cup I(P^I_u+j)}\setminus {i}}$ are also accessible to user $u$, since $u\in I(P^I_S+j)=\{u\}\cup T$. Consequently, user $u$ can access all the mini-subfiles except $W_{d_{u}, S, T}$, ensuring successful decoding of its requested mini-subfile from a transmission of this type.
	\item\textit{General Case}: Consider the user $u$ within a transmission as described in Line \ref{generalcase} of Algorithm \ref{Algo1}. Each mini-subfile within the transmission, except $W_{d_{u},S,T}$, follows the structure $\{W_{d_{I(P^I_u+i)},I(P^I_S+i),I(P^I_T+i)}:I(P^I_S+i)\text{ is a subfile-index set}\}$ for some $i\in[\gamma_aL+\gamma_p]$. Since $1\leq i\leq \gamma_aL+\gamma_p$, we infer $I(P^I_u+i)\neq u$ for any $i$. However, we know that, $I(P^I_u+i)\cup I(P^I_S+i)\cup I(P^I_T+i)=I$. Consequently, $u\in I(P^I_S)$ or $u\in I(P^I_T)$. Thus, user $u$ can access all mini-subfiles in the transmission except $W_{d_{u}, S, T}$, thereby obtaining its desired mini-subfile.
\end{enumerate}
Having explained the decodability of the proposed scheme, we now prove that in the absence of access caches, the above scheme reduces to the well-known MAN scheme\cite{MAN}. 
\begin{remark}
	Consider a $(K, L, M_a=0, M_p, N)-$CW-MAP coded caching system, where access caches have no memory and users solely rely on the contents of their private caches. This scenario mirrors the setting explored in the MAN scheme\cite{MAN}. In this CW-MAP coded caching system, the demanded mini-subfiles follow the structure $W_{d_u,\emptyset, T}$ since $\gamma_a=\frac{KM_a}{N}=0$. With $\gamma_aL=0$, the contents of the private cache of user $u$ is $\mathcal{Z}_u^p=\{W_{n,\emptyset,T}:T\in\binom{[K]\setminus\emptyset}{\gamma_p},u\in T,\forall n\in [N]\}=\{W_{n,\emptyset,T}:T\in\binom{[K]}{\gamma_p},u\in T,\forall n\in [N]\}$. Consequently, the subpacketization for this CW-MAP system equals $F=\binom{K}{\gamma_p}$, which is equal to the subpacketization of the MAN scheme\cite{MAN} for $t=\gamma_p$. For the mini-subfile $W_{d_u,\emptyset,T}$, we have $I=\{u\}\cup T$. There is just one set with $\gamma_aL=0$ consecutive integers in $I$, which is $S=\emptyset$. Thus, all transmissions follow the \textit{Special Case 1}. Consequently, the transmission corresponding to mini-subfile $W_{d_u,\emptyset,T}$ has the form $W_{d_u,\emptyset,T}\bigoplus\limits_{j=1}^{\gamma_p} W_{d_{T_j},\emptyset,\{\{u\}\cup T\}\setminus T_j}$, where $T_j$ represents the $j^{th}$ element of set $T$, ordered lexicographically. Here, we fix an element $i$ of the set $I$ and assign $I\setminus\{i\}$ in place of the mini-subfile-index set. Note that $|I|=\gamma_p+1$, so $|I\setminus\{i\}|=\gamma_p$. This is done for all $i\in I$, and the mini-subfiles are XORed. This is the same delivery scheme as the MAN scheme\cite{MAN}.
\end{remark}

\subsection{Rate-Memory Trade-off Characterization for the Proposed Scheme}
We characterize the rate performance of the scheme proposed in Section \ref{achievability} in this subsection. Note that while the proposed scheme works for any general $\gamma_a$ and $\gamma_p$, the performance characterization is done for the memory regime where $\gamma_p<\gamma_aL$. We will now explain how the rate for the proposed scheme is calculated. 

For a given mini-subfile $W_{d_u,S,T}$, the set $I=\{u\}\cup S \cup T$, which is of cardinality $(1+\gamma_aL+\gamma_p)$, is called its union set. Consider a set $I \subseteq [K]$ of cardinality $(1+\gamma_aL+\gamma_p)$ containing $i$ subfile-index sets. Each subfile-index set $S \subset I$, of cardinality $\gamma_aL$, offers $\binom{1+\gamma_p}{\gamma_p}=(1+\gamma_p)$ ways for selecting a mini-subfile-index set from $I$. From a given set $I$ of cardinality $(1+\gamma_aL+\gamma_p)$, once a subfile-index set $S$ of cardinality $\gamma_aL$ and a mini-subfile-index set of cardinality $\gamma_p$ are chosen, then there is only one way of choosing the user-index from $I$, which is given by $I \setminus \{S \cup T\}$. Thus, for a given set $I$ of cardinality $(1+\gamma_aL+\gamma_p)$ and a chosen subfile-index set $S \subset I$, there are $(1+\gamma_p)$ mini-subfiles, each with the same subfile-index set $S$, and the same union set $I$. Therefore, for a set $I$ of cardinality $(1+\gamma_aL+\gamma_p)$ containing $i$ subfile-index sets, there are a total of $i(1+\gamma_p)$ mini-subfiles, each having the union set as $I$.

 Moving forward, a subset of $[K]$ of cardinality $(1+\gamma_aL+\gamma_p)$ having at least one subfile-index set will be referred to as a \textit{transmission-subset}. Observe that, corresponding to the \textit{General Case} of Algorithm \ref{Algo1}, for a \textit{transmission-subset} $I$ having $i \geq 2$ subfile-index sets, the transmission constructed is a coded combination of $i$ mini-subfiles, as described in Line \ref{generalcase} of Algorithm \ref{Algo1}. Since there are $i(1+\gamma_p)$ mini-subfiles for $I$, the server makes $(1+\gamma_p)$ transmissions for $I$ corresponding to the \textit{General Case}. Let $C_{GC}$ be the number of \textit{transmission-subsets} and $X_{GC}$ be the number of transmissions, respectively, for the \textit{General Case}. Then, we have $X_{GC}=(1+\gamma_p)C_{GC}$. 

We now address \textit{Special Case 1}. Consider a \textit{transmission-subset} $I$ having only one subfile-index set. Hence, there are $(1+\gamma_p)$ mini-subfiles corresponding to this \textit{transmission-subset}. Note that the transmission corresponding to \textit{Special Case 1}, as described in Line \ref{specialcase1}, is a coded combination of $(1+\gamma_p)$ mini-subfiles. Thus, unlike the \textit{General Case}, the server makes just one transmission for a \textit{transmission-subset} in \textit{Special Case 1} instead of $(1+\gamma_p)$ transmissions. In other words, let $C_{SC1}$ be the number of \textit{transmission-subsets} and $X_{SC1}$ be the number of transmissions for the \textit{Special Case 1}, respectively. Then, we have $X_{SC1}=C_{SC1}$.

Finally, consider \textit{Special Case 2}. Consider a \textit{transmission-subset} $I$ that contains only two subfile-index sets with no intersection between them. Since there are two subfile-index sets in $I$, there are $2(1+\gamma_p)$ mini-subfiles with the corresponding union set as $I$. However, the transmission constructed in Line \ref{specialcase2} is a coded combination of these $2(1+\gamma_p)$ mini-subfiles. Hence, the server, similar to \textit{Special Case 1}, makes only one transmission for each such \textit{transmission-subset} instead of $(1+\gamma_p)$ transmissions. Thus, if $C_{SC2}$ denotes the number of \textit{transmission-subsets} and $X_{SC2}$ the number of transmissions for \textit{Special Case 2}, respectively, then, we have $X_{SC2}=C_{SC2}$. 

Let $X$ be the total number of transmissions returned by Algorithm \ref{Algo1} and $C=C_{GC}+C_{SC1}+C_{SC2}$ be the total number of \textit{transmission-subsets}. Then, we have
\begin{align}
	X&=X_{GC}+X_{SC1}+X_{SC2}\nonumber\\
	&=(1+\gamma_p)C_{GC}+C_{SC1}+C_{SC2}\nonumber\\
	&=(1+\gamma_p)C-\gamma_p(C_{SC1}+C_{SC2}).
	\label{eq:nTx}
\end{align}The values of $C, C_{SC1}$ and $C_{SC2}$ have been summarized in Table \ref{table1}. Hence, to calculate the total number of transmissions, we first compute $(1+\gamma_p)C$ and then subtract $\gamma_p(C_{CS1}+C_{CS2})$ from it. Finally, to calculate the rate, we divide the expression obtained above by the subpacketization $F=K\binom{K-\gamma_aL}{\gamma_p}$.
\begin{table*}[t]
	\caption{Total number of \textit{transmission-subsets} and \textit{transmission-subsets} corresponding to \textit{Special Case 1} and \textit{Special Case 2} for $\gamma_p<\gamma_aL-1$ and $\gamma_p=\gamma_aL-1$.}
	\begin{center}
		\begin{tabular}{|c|c|c|}
			\hline
			{} &$\gamma_p<\gamma_aL-1$ &$\gamma_p=\gamma_aL-1$\\
			\hline
			$C$ &\makecell{$\binom{K-\gamma_aL}{1+\gamma_p}+(K-1)\binom{K-\gamma_aL-1}{1+\gamma_p}$\\$-\sum\limits_{i=1}^{\gamma_aL-1} \binom{K-2\gamma_aL-1+i}{1+\gamma_p-\gamma_aL+i}$} &\makecell{$\binom{K-\gamma_aL}{1+\gamma_p}+(K-1)\binom{K-\gamma_aL-1}{1+\gamma_p}-\left(\frac{(K-2\gamma_aL)(K-2\gamma_aL+1)}{2}\right)^+-$\\$((\gamma_aL-1)(K-2\gamma_aL-1))^+-\sum\limits_{i=1}^{\gamma_aL-1}\binom{K-2\gamma_aL-1+i}{1+\gamma_p-\gamma_aL+i}$} \\
			\hline
			$C_{SC1}$ &$K\binom{K-\gamma_aL-2}{1+\gamma_p}$ &\makecell{$K\binom{K-\gamma_aL-2}{1+\gamma_p}-K(K-2\gamma_aL-1)^+$}\\
			\hline
			$C_{SC2}$ &0 &$(1+\gamma_aL)(K-2\gamma_aL-1)+\frac{(K-2\gamma_aL-2)(K-2\gamma_aL-1)}{2}$ \\
			\hline
		\end{tabular}
	\end{center}
	\label{table1}
\end{table*}

We now explain the computation of the total number of transmission subsets, $C$, the number of transmission subsets corresponding to \textit{Special Case 1}, $C_{SC1}$ and the number of transmission subsets corresponding to \textit{Special Case 1}, $C_{SC2}$. Once we obtain these values, the total number of transmissions made can be computed as given in \eqref{eq:nTx} and the rate of transmission is obtained by normalizing the total number of transmissions by the subpacketization $F=K\binom{K-\gamma_aL}{\gamma_p}$.

%We explain the rate calculation below. First, we describe the rate calculation for the \textit{General Case}, as outlined in Line \ref{generalcase} of Algorithm \ref{Algo1}, ignoring \textit{Special Case 1} and \textit{Special Case 2}. In other words, we compute $\frac{(1+\gamma_p)C}{K\binom{K-\gamma_aL}{\gamma_p}}$. Next, we address the rate calculation for \textit{Special Case 1}, detailed in Line \ref{specialcase1}. Finally, we explain the rate calculation for \textit{Special Case 2}, as specified in Line \ref{specialcase2}, calculating the rate expression $\frac{(1+\gamma_p)C-\gamma_p(C_{SC1}+C_{SC2})}{K\binom{K-\gamma_aL}{\gamma_p}}$.
\begin{itemize}
	\item \textit{Total number of transmission sets $C$}: We calculate the total number of \textit{transmission-subsets} $C$ for $\gamma_p<\gamma_aL-1$ and for $\gamma_p=\gamma_aL-1$ separately as follows. 
	
	{When $\gamma_p<\gamma_aL-1$}: Let us start by considering the subfile-index set $S = [\gamma_aL]$. To form a \textit{transmission-subset}, we need to choose $(1+\gamma_p)$ elements from the set $[K]$ excluding those already present in the subfile-index set $S$. Thus, the number of \textit{transmission-subsets} containing the subfile-index set $S =[\gamma_aL]$ is $\binom{K-\gamma_aL}{1+\gamma_p}$. 
	
	Next, let us examine the subfile-index set $[2,\gamma_aL+1]$. Any selection of $(1+\gamma_p)$ elements containing the element $1$ would result in a \textit{transmission-subset} already counted for the subfile-index set $[\gamma_aL]$. Therefore, we choose $(1+\gamma_p)$ elements in the set $[K]$, excluding the elements from $[2,\gamma_aL+1]$ and the element $1$. Thus, the number of \textit{transmission-subsets} corresponding to the subfile-index set $[2,\gamma_aL+1]$ is $\binom{K-1-\gamma_aL}{1+\gamma_p}$. This pattern holds for all subfile-index sets $[i,i+\gamma_aL-1]$ for $2\leq i\leq K-\gamma_aL+1$. For a subfile-index set $[i,i+\gamma_aL-1]$, a subset of size $(1+\gamma_p)$ can be chosen from the set $[K] \setminus [i-1, i+\gamma_aL-1]$, for $2\leq i\leq K-\gamma_aL+1$.

	Now, we explain the calculation of the number of \textit{transmission-subsets} associated with subfile-index sets $<[i,i+\gamma_aL-1]>_K$ for $i\in[K-\gamma_aL+2, K]$. Consider the subfile-index set $<[K-\gamma_aL+2,K+1]>_K$. As explained earlier, we select $(1+\gamma_p)$ elements from the set $[K]$ excluding the elements in the set $<[K-\gamma_aL+2,K+1]>_K$ and the element $K-\gamma_aL+1$. Thus, we can choose $(1+\gamma_p)$ elements from the set of elements $\{2,3,\cdots, K-\gamma_aL\}$, among which $2,3,\cdots, \gamma_aL$, are part of the subfile-index set $[\gamma_aL]$. If these elements are included in the $(1+\gamma_p)$ elements selected, it would lead to double counting with the \textit{transmission-subsets} associated with the subfile-index set $[\gamma_aL]$. Consequently, we must ensure that the set $[2,\gamma_aL]$ is not present among the selected elements. Hence, the number of \textit{transmission-subsets} for the subfile-index set $<[K-\gamma_aL+2,K+1]>_K$ is $\binom{K-\gamma_aL-1}{1+\gamma_p}-\binom{K-\gamma_aL-1-(\gamma_aL-1)}{1+\gamma_p-(\gamma_aL-1)}$.

	Next, consider the subfile-index set $<[K-\gamma_aL+3,K+2]>_K$. For this subfile-index set, we can choose $(1+\gamma_p)$ elements from the set $[K]$ excluding the elements in the set $<[K-\gamma_aL+3,K+2]>_K$ and the element $K-\gamma_aL+2$, that is, we can choose $(1+\gamma_p)$ elements from the set $\{3,4,\cdots,K-\gamma_aL+1\}$. However, observe that if the $(1+\gamma_p)$ selected elements include the set $[3,\gamma_aL]$, it would lead to double counting with the \textit{transmission-subsets} associated with the subfile-index set $[\gamma_aL]$. Further, if the $(1+\gamma_p)$ selected elements include the set $[3,\gamma_aL+1]$, it would lead to double counting with the \textit{transmission-subsets} associated with the subfile-index set $[\gamma_aL]$, as well as, with the \textit{transmission-subsets} associated with the subfile-index set $[2,\gamma_aL+1]$. Therefore, to avoid any double counting of the \textit{transmission-subsets}, we have to remove any \textit{transmission-subset} containing the set $[3,\gamma_aL]$. Thus, the number of \textit{transmission-subsets} for the subfile-index set $<[K-\gamma_aL+3,K+2]>_K$ is $\binom{K-\gamma_aL-1}{1+\gamma_p}-\binom{K-\gamma_aL-1-(\gamma_aL-2)}{1+\gamma_p-(\gamma_aL-2)}$. Following the above logic, the number of \textit{transmission-subsets} for the subfile-index set $<[i,i+\gamma_aL-1]>_K$ for some $i\in[K-\gamma_aL+1,K]$, that is, for the subfile-index set $<[K-\gamma_aL+1+i,K+i]>_K$ is $\binom{K-\gamma_aL-1}{1+\gamma_p}-\binom{K-\gamma_aL-1-(\gamma_aL-i)}{1+\gamma_p-(\gamma_aL-i)}$, for $i\in[\gamma_aL-1]$.
	
	Thus, the total number of \textit{transmission-subsets} when $\gamma_p<\gamma_aL-1$ is $C = \binom{K-\gamma_aL}{1+\gamma_p}+(K-1)\binom{K-\gamma_aL-1}{1+\gamma_p}-\sum\limits_{i=1}^{\gamma_aL-i} \binom{K-\gamma_aL-1-(\gamma_aL-i)}{1+\gamma_p-(\gamma_aL-i)}$.

{\textit{Number of \textit{transmission-subsets} for $\gamma_p=\gamma_aL-1$}}: Since $\gamma_p=\gamma_aL-1$, selecting $(1+\gamma_p)=(1+\gamma_aL-1)=\gamma_aL$ elements for the \textit{transmission-subsets} can result in selecting a subfile-index set that has previously occurred. Thus, we will count the number of such subfile-index sets and remove them from the \textit{transmission-subset} calculation. %We illustrate this point using the following example.
Notice that the calculation of the number of \textit{transmission-subsets} starts by considering the subfile-index set $[\gamma_aL]$. Then, we calculate the \textit{transmission-subsets} for the subfile-index set $[2,\gamma_aL+1]$ and so on. Therefore, we only need to check for subfile-index sets that have been considered before the current subfile-index set. Consider the subfile-index set $<[\gamma_aL+2,2\gamma_aL+1]>_K$. When selecting $(1+\gamma_p)$ elements for this subfile-index set, we remove the element $\gamma_aL+1$ from consideration along with the elements in the set $<[\gamma_aL+2,2\gamma_aL+1]>_K$, as explained earlier. Thus, we select $(1+\gamma_p)$ elements from the set $\{1,2,\cdots,\gamma_aL,2\gamma_aL+2,\cdots,K\}$. However, we can still select the set $<[\gamma_aL]>_K$, resulting in double counting. Similarly, for the set $<[\gamma_aL+3,2\gamma_aL+2]$, we pick $(1+\gamma_p)$ elements from the set $\{1,2,\cdots,\gamma_aL+1,2\gamma_aL+3,\cdots,K\}$ and double counting can occur if we select %$(1+\gamma_p)$ elements from 
either the subfile-index set $<[\gamma_aL]>_K$ or the subfile-index set $<[2,\gamma_aL+1]>_K$.

Thus, for the subfile-index set 
$<[i,i+\gamma_aL-1]>_K$ for $i\in[\gamma_aL+2,K-\gamma_aL+1]$, double counting occurs if we select a subfile-index set from the set $<[i-2]>_K$. Since, there are $(i-1-\gamma_aL)$ subfile-index sets in the set $<[i-2]>_K$, summing over $i\in[\gamma_aL+2,K-\gamma_aL+1]$, we have $\frac{(K-2\gamma_aL)(K-2\gamma_aL+1)}{2}$ \textit{transmission-subsets} that will be counted more than once. 

We now explain the procedure for determining the number of \textit{transmission-subsets} associated with the subfile-index sets $<[i,i+\gamma_aL-1]>_K$ for $i\in[K-\gamma_aL+2,K]$. Towards this end, consider the subfile-index set $<[K-\gamma_aL+2,K+1]>_K$. For this subfile-index set, we can choose $(1+\gamma_p)$ elements from the set $\{2,3,\cdots,K-\gamma_aL\}$. However, we have already removed \textit{transmission-subsets} containing the set $[2,\gamma_aL]$ from consideration, as explained before. Hence, we can not consider the subfile-index set $[2,\gamma_aL+1]$ here. Thus, we need to remove the subfile-index sets which have been considered before from the set $\{2,3,\cdots,K-\gamma_aL\}$. There are $(K-2\gamma_aL-1)$ such subfile-index sets. Similarly, for the subfile-index set $<[K-\gamma_aL+3,K+2]>_K$, we need to check for subfile-index sets which have been considered before from the set $\{3,4,\cdots, K-\gamma_aL+1\}$. Again, there are $(K-2\gamma_aL-1)$ such subfile-index sets.

Since there are $(K-2\gamma_aL-1)$ subfile-index sets in the set $<[i-(K-\gamma_aL),i-2]>_K$ and there are $\gamma_aL-1$ subfile-index set of the form $<[i,i+\gamma_aL-1]>_K,i\in[K-\gamma_aL+2,K]$, we have $(\gamma_aL-1)(K-2\gamma_aL-1)$ \textit{transmission-subsets} that will be counted more than once. Using the above calculations, we obtain the total number of transmission subsets as $C=(1+\gamma_p)\bigg(\binom{K-\gamma_aL}{1+\gamma_p}+(K-1)\binom{K-\gamma_aL-1}{1+\gamma_p}-\left(\frac{(K-2\gamma_aL)(K-2\gamma_aL+1)}{2}\right)^+-((\gamma_aL-1)(K-2\gamma_aL-1))^+-\sum\limits_{i=1}^{\gamma_aL-1}\binom{K-\gamma_aL-1-(\gamma_aL-i)}{1+\gamma_p-(\gamma_aL-i)}\bigg)$.

	\item \textit{Special Case 1}: We will now describe the rate calculations for \textit{Special Case 1}.
	
	{\textit{Number of \textit{transmission-subsets} for $\gamma_p<\gamma_aL-1$}}: Consider a \textit{transmission-subset} with only one subfile-index set, $<[i,i+\gamma_aL-1]>_K$ for some $i\in[ K]$. We need to select the remaining $(1+\gamma_p)$ elements so that no other subfile-index sets are possible. Note that the elements $<i-1>_K$ and $<i+\gamma_aL>_K$ can not be selected since they will lead to the presence of subfile-index sets $<[i-1,i+\gamma_aL-2]>_K$ and $<[i+1,i+\gamma_aL]>_K$, respectively, in the \textit{transmission-subsets}. Since, $(1+\gamma_p)<\gamma_aL$, no additional subfile-index sets will be formed on choosing $(1+\gamma_p)$ elements from the remaining $(K-\gamma_aL-2)$ elements. Since there are $\binom{K-\gamma_aL-2}{1+\gamma_p}$ \textit{transmission-subsets} for each subfile-index set of the form $<[i,i+\gamma_aL-1]>_K,i\in[K]$, there are $K\binom{K-\gamma_aL-2}{1+\gamma_p}$ total \textit{transmission-subsets}. The server makes one transmission for each such \textit{transmission-subset} instead of $(1+\gamma_p)$ transmissions, and thus there is a reduction in the number of transmissions by $\gamma_pK\binom{K-\gamma_aL-2}{1+\gamma_p}$ as reflected in \eqref{equationrate}.
	
	{\textit{Number of \textit{transmission-subsets} for $\gamma_p=\gamma_aL-1$}}: For this case, we have $(1+\gamma_p)=\gamma_aL$. Therefore, we need to remove any subfile-index sets that can be chosen from the $(K-\gamma_aL-2)$ elements. However, since the subfile-index sets comprise consecutive integers, the $(K-\gamma_aL-2)$ elements also comprise consecutive integers in a cyclic wrap-around fashion. For example, consider the subfile-index set $<[i,i+\gamma_aL-1]$. As explained before, for this subfile-index set, we can select $(1+\gamma_p)$ elements from the set $[K]\setminus <[i-1,i+\gamma_aL]>_K$, that is, from the set $\{i+\gamma_aL+1,\cdots,K,1,\cdots,i-2\}$. Notice that the elements of the set $\{i+\gamma_aL+1,\cdots,K,1,\cdots,i-2\}$ are consecutive in a cyclic wrap-around sense. Because there are $(K-2\gamma_aL-1)$ subfile-index sets in $(K-\gamma_aL-2)$ consecutive integers, we have a total reduction of $\gamma_p K\left(\binom{K-\gamma_aL-2}{1+\gamma_p}-(K-2\gamma_aL-1)^+\right)$ transmissions, as shown in \eqref{equationrate}.
	\item \textit{Special Case 2}: 
	We now explain the rate calculation for \textit{Special Case 2}. 
	Consider a subfile-index set $<[i,i+\gamma_aL-1]>_K$. Selecting either $<i-1>_K$ or $<i+\gamma_aL>_K$ will lead to a \textit{transmission-subset} having two subfile-index sets. However, these subfile-index sets will have an intersection between them. Thus, we cannot select the elements $<i-1>_K$ or $<i+\gamma_aL>_K$.

	Consider the subfile-index set $[\gamma_aL]$. For this subfile-index set, we select subfile-index sets from the set $[K]\setminus \{K,1,\cdots,\gamma_aL+1\}$, that is, the set $[\gamma_aL+2,K-1]$ to form \textit{transmission-subsets}. Thus, there are $K-\gamma_aL-2$ elements from which we choose subsets of size ($1+\gamma_p) = \gamma_aL$ containing consecutive elements as $[j,j+\gamma_aL-1], j \in [\gamma_aL+2,K-\gamma_aL]$. Therefore, we have a total of $K-2\gamma_aL-1$ \textit{transmission-subsets}. Similarly, for each of the subfile index sets $[2,\gamma_aL+1], [3,\gamma_aL+2], \cdots, <[\gamma_aL+1,2\gamma_aL]>_K$ also, we can choose $K-2\gamma_aL-1$ \textit{transmission-subsets}. 
	
	Now, consider the subfile-index set $<[\gamma_aL+2,2\gamma_aL+1]>_K$. For this subfile-index set, we choose subfile-index sets from the set $\{1,2,\cdots,\gamma_aL,2\gamma_aL+3,\cdots,K\}$ to form \textit{transmission-subsets}. However, since we have already formed \textit{transmission-subsets} for the subfile-index set $[\gamma_aL]$, we can not select the set $[\gamma_aL]$ to form a \textit{transmission-subset} for the subfile-index set $<[\gamma_aL+2,2\gamma_aL+1]>_K$. Thus, there are $K-2\gamma_aL-2$ \textit{transmission-subsets} for the subfile-index set $<[\gamma_aL+2,2\gamma_aL+1]>_K$. Similarly, for the subfile-index set $<[\gamma_aL+3,2\gamma_aL+2]>_K$, we choose subfile-index sets from the set $\{1,2,\cdots,\gamma_aL+1,2\gamma_aL+4,\cdots,K\}$. Since, the subfile-index sets $[\gamma_aL]$ and $[2,\gamma_aL+1]$ have already been considered, there are $K-2\gamma_aL-3$ \textit{transmission-subsets} for the subfile-index set $<[\gamma_aL+3,2\gamma_aL+2]>_K$.

	Thus, as $i$ increases past $\gamma_aL+2$, the number of $(1+\gamma_p)$ sets we can select keeps on decreasing by one. 
	Hence, there are a total of $(1+\gamma_aL)(K-2\gamma_aL-1)+(K-2\gamma_aL-2)+(K-2\gamma_aL-3)+\cdots+2+1%1+2+\cdots+K-2\gamma_aL-2
	=(1+\gamma_aL)(K-2\gamma_aL-1)+\frac{(K-2\gamma_aL-2)(K-2\gamma_aL-1)}{2}$ such \textit{transmission-subsets}, resulting in a reduction of $\gamma_p\bigg((1+\gamma_aL)(K-2\gamma_aL-1)+\frac{(K-2\gamma_aL-2)(K-2\gamma_aL-1)}{2}\bigg)$. This is reflected in \eqref{equationrate}.

\end{itemize}

\subsection{Memory Sharing to Achieve Rate for Non-Integral $\gamma_a$ and $\gamma_p$}
\label{memorysharing}
The rate $R_{M_a,M_p}$ is defined only for integer values of $\gamma_a$ and $\gamma_p$ in Theorem \ref{thm2}. However, the rate points for non-integral values of $\gamma_a$ and $\gamma_p$ can be achieved via memory sharing. We address the case where $\gamma_a\not\in\mathbb{Z}^+$ or $\gamma_p\not\in\mathbb{Z}^+$ below. 

Consider the case where $\gamma_a\not\in\mathbb{Z}^+$ and $\gamma_p\in\mathbb{Z}^+$. Define $M_{a_1}\triangleq\frac{N\lfloor\gamma_a\rfloor}{K}$ and $M_{a_2}\triangleq\frac{N\lceil\gamma_a\rceil}{K}$. Since $M_{a_1}\leq M_a\leq M_{a_2}$, we can write $M_a$ as $M_a=\alpha_1M_{a_{1}}+(1-\alpha_1)M_{a_{2}}$, for some $0\leq \alpha_1\leq 1$. Each file $W_{n}$ is broken down as $W_{n}=\{W^{\alpha_1}_{n}, W^{1-\alpha_1}_{n}\}$ where $|W^{\alpha_1}_{n}|=\alpha_1B$ bits and $|W^{1-\alpha_1}_{n}|=(1-\alpha_1)B$ bits. The files $W_{n}^{\alpha_1}$ and $W_{n}^{1-\alpha_1}$ are broken down into subfiles as $W^{\alpha_1}_{n}=\{W^{\alpha_1}_{n,i}:i\in[K]\}$ and $W^{1-\alpha_1}_{n}=\{W^{1-\alpha_1}_{n,i}:i\in[K]\}$ in accordance with \eqref{subfileconstruction} and placed in the access caches as described by \eqref{placementaccesscaches} for $\lfloor\gamma_a\rfloor$ and $\lceil\gamma_a\rceil,$ respectively. Hence, every access cache stores $\frac{N\alpha_1\lfloor\gamma_a\rfloor B+N(1-\alpha_1)\lceil\gamma_a\rceil B}{K}=\alpha_1M_{a_{1}}B+(1-\alpha_1)M_{a_{2}}B=M_aB$ bits, satisfying its memory constraint. 

The subfiles $W_{n,i}^{\alpha_1}$ and $W_{n,i}^{1-\alpha_1}$ are broken down into mini-subfiles as dictated by \eqref{minisubfileconstruction} and the private cache of the users is populated with these mini-subfiles as described in \eqref{placementpolicyprivatecaches} for $\lceil \gamma_a\rceil$ and $\lfloor \gamma_a\rfloor$, respectively. Every private cache stores $\alpha_1\frac{N(K-\lfloor\gamma_a\rfloor L)\binom{K-\lfloor\gamma_a\rfloor L-1}{\gamma_p-1}}{K\binom{K-\lfloor\gamma_a\rfloor L}{\gamma_p}}+(1-\alpha_1)\frac{N(K-\lceil\gamma_a\rceil L)\binom{K-\lceil\gamma_a\rceil L-1}{\gamma_p-1}}{K\binom{K-\lceil\gamma_a\rceil L}{\gamma_p}}=\alpha_1\frac{N\gamma_p}{K}+(1-\alpha_1)\frac{N\gamma_p}{K}=M_p$ files, satisfying its memory constraint.

Finally, the rate $R_{M_a,M_p}$ is given as 
\begin{align}
	R_{M_a,M_p}=\alpha_1R_{M_{a_{1}},M_p}+(1-\alpha_1)R_{M_{a_{2}},M_p}.
\end{align} 
Next, we address the case where $\gamma_a\in\mathbb{Z}^+$ and $\gamma_p\not\in\mathbb{Z}^+$. Define $M_{p_{1}}\triangleq\frac{N\lfloor\gamma_p\rfloor}{K}$ and $M_{p_{2}}\triangleq\frac{N\lceil\gamma_p\rceil}{K}$. We can write $M_p$ as $M_p=\alpha_2M_{p_{1}}+(1-\alpha_2)M_{p_{2}}$, for some $0\leq \alpha_2\leq 1$, as $M_{p_1}\leq M_p\leq M_{p_2}$. Each file $W_{n}$ is split into subfile $W_{n,S}$ as described in \eqref{subfileconstruction}. Each subfile $W_{n,S}$ is further split as $W_{n,S}=\{W^{\alpha_2}_{n,S}, W^{1-\alpha_2}_{n,S}\}$, where $|W^{\alpha_2}_{n,S}|=\alpha_2 B_s$ and $|W^{1-\alpha_2}_{n,S}|=(1-\alpha_2) B_s$, where $B_s=\frac{B}{K}$, is the size of one subfile. The subfiles $W^{\alpha_2}_{n,S}$ and $W^{1-\alpha_2}_{n,S}$ are broken down as $W^{\alpha_2}_{n,S}=\{W^{\alpha_2}_{n,S,T}:T\in\binom{[K]\setminus S}{\lfloor\gamma_p\rfloor}\}$ and $W^{1-\alpha_2}_{n,S}=\{W^{1-\alpha_2}_{n,S,T}:T\in\binom{[K]\setminus S}{\lceil\gamma_p\rceil}\}$ as described in \eqref{minisubfileconstruction} and placed in the private caches of the users as described in \eqref{placementpolicyprivatecaches} for $\lfloor\gamma_p\rfloor$ and $\lceil\gamma_p\rceil$, respectively. Thus, each private cache stores $\alpha_2\frac{N(K-\gamma_aL)\binom{K-\gamma_aL-1}{\lfloor\gamma_p\rfloor-1}}{K\binom{K-\gamma_aL}{\lfloor\gamma_p\rfloor}}B+(1-\alpha_2)\frac{N(K-\gamma_aL)\binom{K-\gamma_aL-1}{\lceil\gamma_p\rceil-1}}{K\binom{K-\gamma_aL}{\lceil\gamma_p\rceil}}B=\alpha_2\frac{N\lfloor\gamma_p\rfloor}{K}B+(1-\alpha_2)\frac{N\lceil\gamma_p\rceil}{K}B=\alpha_2M_{p_{1}}B+(1-\alpha_2)M_{p_{2}}B=M_pB$ bits, satisfying its memory constraint. The rate $R_{M_a,M_p}$ is given as 
\begin{align}
	R_{M_a,M_p}=\alpha_2 R_{M_a,M_{p_{1}}}+(1-\alpha_2)R_{M_a,M_{p_{2}}}.
\end{align} 

Finally, we address the case $\gamma_a\not\in\mathbb{Z}^+$ and $\gamma_p\not\in\mathbb{Z}^+$. Define $M_{a_3}\triangleq\frac{N\lfloor\gamma_a\rfloor}{K}$ and $M_{a_4}\triangleq\frac{N\lceil\gamma_a\rceil}{K}$. Since $M_{a_3}\leq M_a\leq M_{a_4}$, we can write $M_a$ as $M_a=\alpha_3M_{a_{3}}+(1-\alpha_3)M_{a_{4}}$, for some $0\leq \alpha_3\leq 1$. Each file $W_{n}$ is broken down as $W_{n}=\{W^{\alpha_3}_{n}, W^{1-\alpha_3}_{n}\}$ where $|W^{\alpha_3}_{n}|=\alpha_3B$ bits and $|W^{1-\alpha_3}_{n}|=(1-\alpha_3)B$ bits. The files $W_{n}^{\alpha_3}$ and $W_{n}^{1-\alpha_3}$ are broken down into subfiles as $W^{\alpha_3}_{n}=\{W^{\alpha_3}_{n,i}:i\in[K]\}$ and $W^{1-\alpha_3}_{n}=\{W^{1-\alpha_3}_{n,i}:i\in[K]\}$ in accordance with \eqref{subfileconstruction} and placed in the access caches as described by \eqref{placementaccesscaches} for $\lfloor\gamma_a\rfloor$ and $\lceil\gamma_a\rceil,$ respectively. Hence, every access cache stores $\frac{N\alpha_3\lfloor\gamma_a\rfloor B+N(1-\alpha_3)\lceil\gamma_a\rceil B}{K}=\alpha_3M_{a_{3}}B+(1-\alpha_3)M_{a_{4}}B=M_aB$ bits, satisfying its memory constraint. 

Next, define $M_{p_{3}}\triangleq\frac{N\lfloor\gamma_p\rfloor}{K}$ and $M_{p_{4}}\triangleq\frac{N\lceil\gamma_p\rceil}{K}$. We can write $M_p$ as $M_p=\alpha_4M_{p_{3}}+(1-\alpha_4)M_{p_{4}}$, for some $0\leq \alpha_4\leq 1$, as $M_{p_3}\leq M_p\leq M_{p_4}$. The subfiles $W^{\alpha_3}_{n,i}$ and $W_{n,i}^{1-\alpha_3}$ are further split as $W^{\alpha_3}_{n,i}=\{W^{\alpha_3,\alpha_4}_{n,i}, W^{\alpha_3,1-\alpha_4}_{n,i}\}$, and $W^{1-\alpha_3}_{n,i}=\{W^{1-\alpha_3,\alpha_4}_{n,i}, W^{1-\alpha_3,1-\alpha_4}_{n,i}\}$, where $|W^{\alpha_3,\alpha_4}_{n,i}|=\alpha_4 B^{\alpha_3}_s$, $|W^{\alpha_3,1-\alpha_4}_{n,i}|=(1-\alpha_4) B^{\alpha_3}_s$, $|W^{1-\alpha_3,\alpha_4}_{n,i}|=\alpha_4 B^{1-\alpha_3}_s$ and $|W^{1-\alpha_3,1-\alpha_4}_{n,i}|=(1-\alpha_3) B^{\alpha_3}_s$, where $B^{\alpha_3}_s=\frac{|W^{\alpha_3}_n|}{K}$, and $B^{1-\alpha_3}_s=\frac{|W^{1-\alpha_3}_n|}{K}$ are the sizes of subfiles $W^{\alpha_3}_{n,i}$ and $W^{1-\alpha_3}_{n,i}$, respectively. These subfiles are broken down as described in \eqref{minisubfileconstruction} and placed in the private caches of the users as described in \eqref{placementpolicyprivatecaches} for $\lfloor\gamma_p\rfloor$ and $\lceil\gamma_p\rceil$, respectively. Hence, each private cache stores $\alpha_3\alpha_4\frac{N(K-\lfloor\gamma_a\rfloor L)\binom{K-\lfloor \gamma_a\rfloor L-1}{\lfloor\gamma_p\rfloor-1}}{K\binom{K-\lfloor\gamma_a\rfloor L}{\lfloor\gamma_p\rfloor}}B+
(1-\alpha_4)\alpha_3\frac{N(K-\lfloor\gamma_a\rfloor L)\binom{K-\lfloor \gamma_a\rfloor L-1}{\lceil\gamma_p\rceil-1}}{K\binom{K-\lfloor\gamma_a\rfloor L}{\lceil\gamma_p\rceil}}B+
(1-\alpha_3)\alpha_4\frac{N(K-\lceil\gamma_a\rceil L)\binom{K-\lceil \gamma_a\rceil L-1}{\lfloor\gamma_p\rfloor-1}}{K\binom{K-\lceil\gamma_a\rceil L}{\lfloor\gamma_p\rfloor}}B+
(1-\alpha_4)(1-\alpha_3)\frac{N(K-\lceil\gamma_a\rceil L)\binom{K-\lceil \gamma_a\rceil L-1}{\lceil\gamma_p\rceil-1}}{K\binom{K-\lceil\gamma_a\rceil L}{\lceil\gamma_p\rceil}}B=\alpha_4M_{p_{3}}B+(1-\alpha_4)M_{p_{4}}B=M_pB$ bits, satisfying its memory constraint.

Finally, the rate $R_{M_a,M_p}$ is given as 
\begin{align}
	&R_{M_a,M_p}=\alpha_3(\alpha_4R_{M_{a_{3}},M_{p_{3}}}+(1-\alpha_4)R_{M_{a_{3}},M_{p_{4}}})+\nonumber\\&(1-\alpha_3)(\alpha_4R_{M_{a_{4}},M_{p_{3}}}+(1-\alpha_4)R_{M_{a_{4}},M_{p_{4}}}).
\end{align} 

%%%%%%%%%%%%%%%%%%%%%%%%%%%%%%%%%%%%%%%%%%%%%%%%%%%%%%%%%%%%%%
\section{Optimality in Large Memory Regime}
\label{ssec:optimality}
In this section, we give the proof of Lemma \ref{lem:opt}. We prove that the proposed scheme is optimal in large memory regimes by showing that the rate of the proposed scheme $R_{M_a,M_p}$ matches the lower bound on the optimal worst-case rate $R^{\textasteriskcentered}_{M_a, M_p}$ derived in Theorem \ref{thm1}, using the cut-set bound\cite{CT}. 
\begin{IEEEproof}
	Since $M_aL+M_p\geq N\left(1-\frac{1}{K}\right)$, we have $\gamma_aL+\gamma_p\geq K-1$. For every sub-file $S$ that a user $u$ does not get on connecting to its respective access caches, user $u$ demands $\binom{K-\gamma_aL-1}{\gamma_p}$ mini-subfiles of $S$. Since every user demands mini-subfiles of $K-\gamma_aL$ subfiles, there will be a total of $K(K-\gamma_aL)\binom{K-\gamma_aL-1}{\gamma_p}$ mini-subfiles that will be demanded by all the users. Hence, there will be either $K(K-\gamma_aL)$ mini-subfiles that are demanded, for $\gamma_p=K-1-\gamma_aL$ or no mini-subfiles that are demanded, for $\gamma_p=K-\gamma_aL$. For the case where $\gamma_p=K-1-\gamma_aL$, the mini-subfiles demanded will be of the form $\{W_{d_u,S,T}:u\notin T,u\not\in S,\text{$S$ is a subfile-index set}, %|S|=\gamma_aL,
	T\in \binom{[K]\setminus S}{K-1-\gamma_aL}\}$. However, for this mini-subfile, we have $I=\{u\}\cup S\cup T=[K]$. Observe that this set has $K$ subfile-index sets. Hence, every transmission will be a coded combination of $K$ mini-subfiles. Therefore, the number of transmissions will be $\frac{K(K-\gamma_aL)}{K}=K-\gamma_aL$ and the rate $R_{M_a,M_p}=\frac{K-\gamma_aL}{K\binom{K-\gamma_aL}{\gamma_p}}=\frac{1}{K}$. For the case where $\gamma_p=K-\gamma_aL$, no mini-subfile is demanded because every user has access to the entire library jointly from the access caches it connects to and its own private cache. For this case, $R_{M_a,M_p}=0$. 
	
	Now, consider the cut-set bound derived in Theorem \ref{thm1}. For $s=1$, we have $R^{\textasteriskcentered}_{M_a,M_p}\geq 1-\frac{M_aL+M_p}{N}\geq 1-\frac{N\left(1-\frac{1}{K}\right)}{N}=\frac{1}{K}$. Since $R_{M_a,M_p}=R^{\textasteriskcentered}_{M_a,M_p}$, the rate of the proposed scheme is optimal given that $M_aL+M_p\geq N\left(1-\frac{1}{K}\right)$. This concludes the proof.
\end{IEEEproof}
While it may seem that the memory regime for which the proposed scheme is proved to be optimal is limited, there are a large number of $(M_a, L, M_p)$ triplets that fall in the optimality regime and each $(M_a, L, M_p)$ triplet corresponds to a different CW-MAP setting. This is also apparent from the numerical plots given in the next section.
\section{Numerical Comparison}
\label{numericalcomparison}
In this section, we compare the rate of the proposed scheme in Theorem \ref{thm2} with the cut-set bound derived in Theorem \ref{thm1}. We provide numerical plots of the rate $R_{M_a, M_p}$ and the lower bound derived in Theorem \ref{thm1} for a system with $K=30$ access caches, $N$ files, and $K=30$ users, such that $N=K$, with access degree $L=3$ and different values of the access cache memory $M_a$ and private cache memory $M_p$. 

\begin{figure}[t]
	\includegraphics[trim={0 0 0 4cm},clip,width=\textwidth]{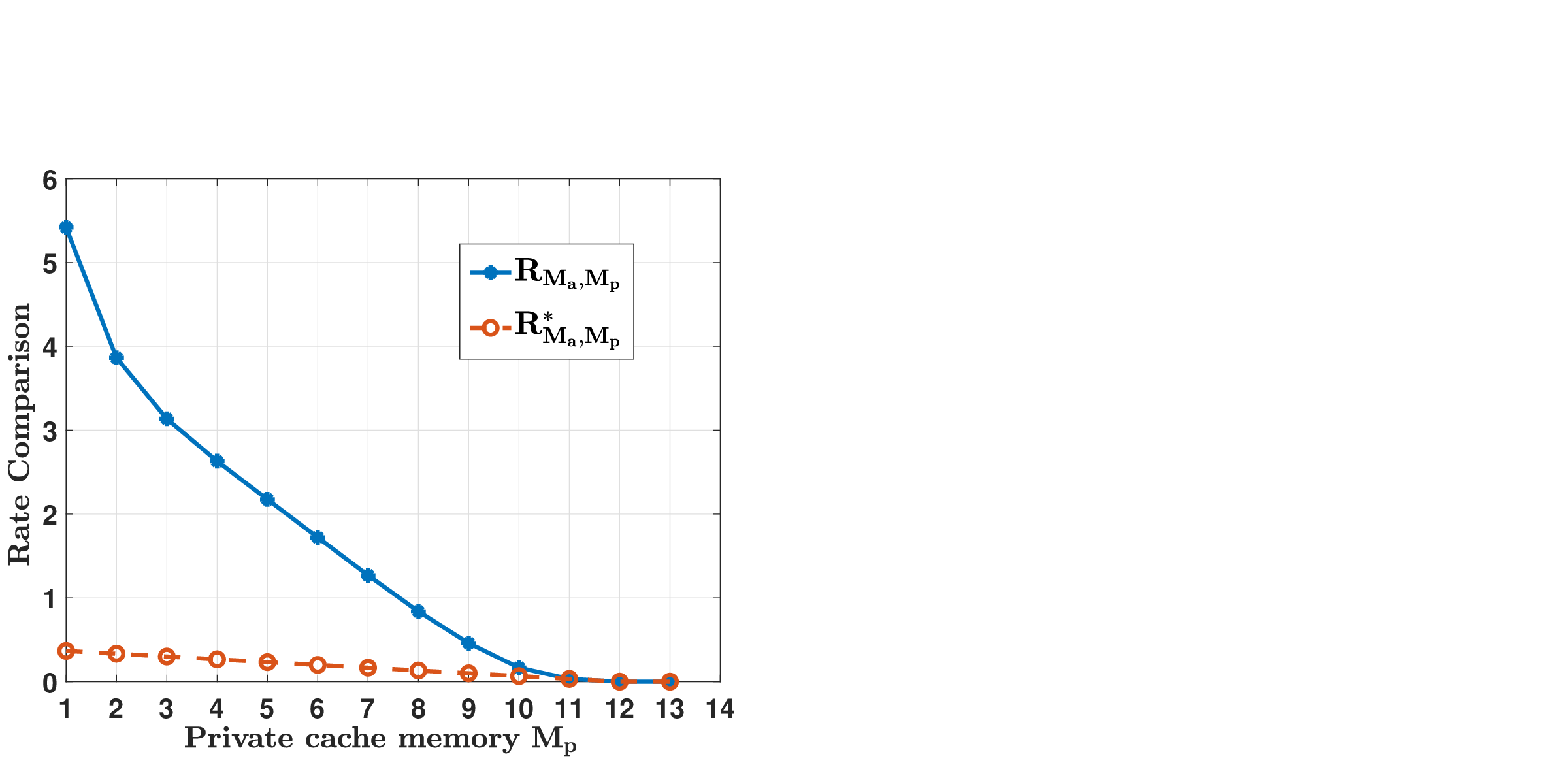}
	\caption{Rate vs $M_p$ comparison for the $(K=30,L=3,M_a=6,M_p,N=30)-$CW-MAP coded caching system.}
	\label{fig2}
	\includegraphics[width=\textwidth]{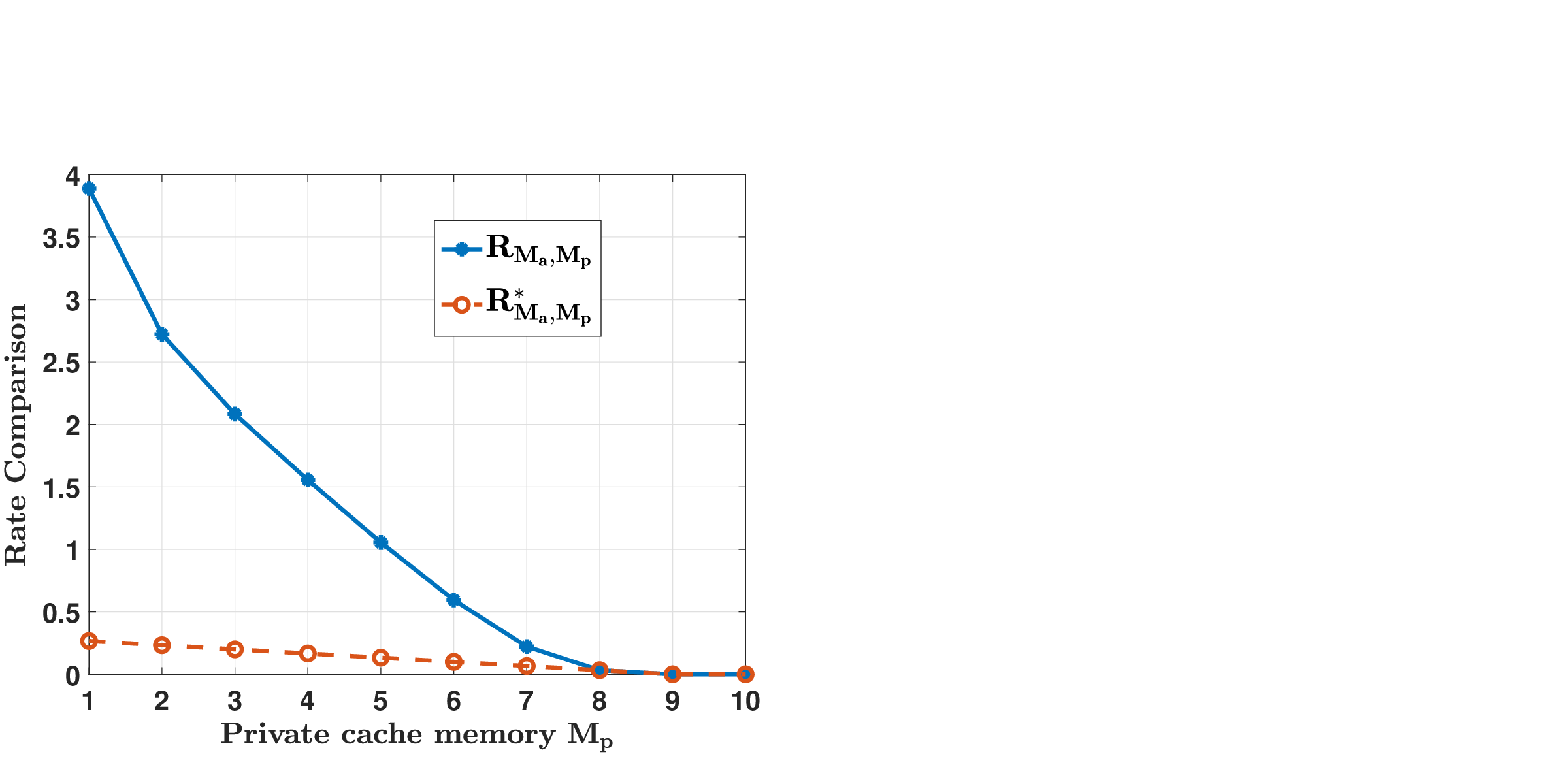}
	\caption{Rate vs $M_p$ comparison for the $(K=30,L=3,M_a=7,M_p,N=30)-$CW-MAP coded caching system.}
	\label{fig3}
\end{figure}

For the CW-MAP coded caching system with $K=30$ access caches, Fig. \ref{fig2}. shows the variation of the rate $R_{M_a, M_p}$ and the lower bound on $R^{\textasteriskcentered}_{M_a, M_p}$ derived in Theorem \ref{thm1} for $L=3$ and $M_a=6$, with $M_p$ taking values from $1$ to $13$. Observe that in Fig. \ref{fig2}., $R_{M_a,M_p}$ moves closer to the lower bound % $R^{\textasteriskcentered}_{M_a,M_p}$ 
as $M_p$ increases, achieving optimality at $M_p=11$.

Fig. \ref{fig3}. shows the variation of the rate $R_{M_a,M_p}$ and the lower bound on $R^{\textasteriskcentered}_{M_a, M_p}$ derived in Theorem \ref{thm1} for $L=3$ and $M_a=7$, with $M_p$ ranging from $1$ to $10$. In Fig. \ref{fig3}., $R_{M_a,M_p}$ approaches the lower bound % $R^{\textasteriskcentered}_{M_a,M_p}$
 as $M_p$ increases, achieving optimality at $M_p=8$.

\begin{figure}[t]
	\includegraphics[trim={0 0 0 4cm},clip,width=\textwidth]{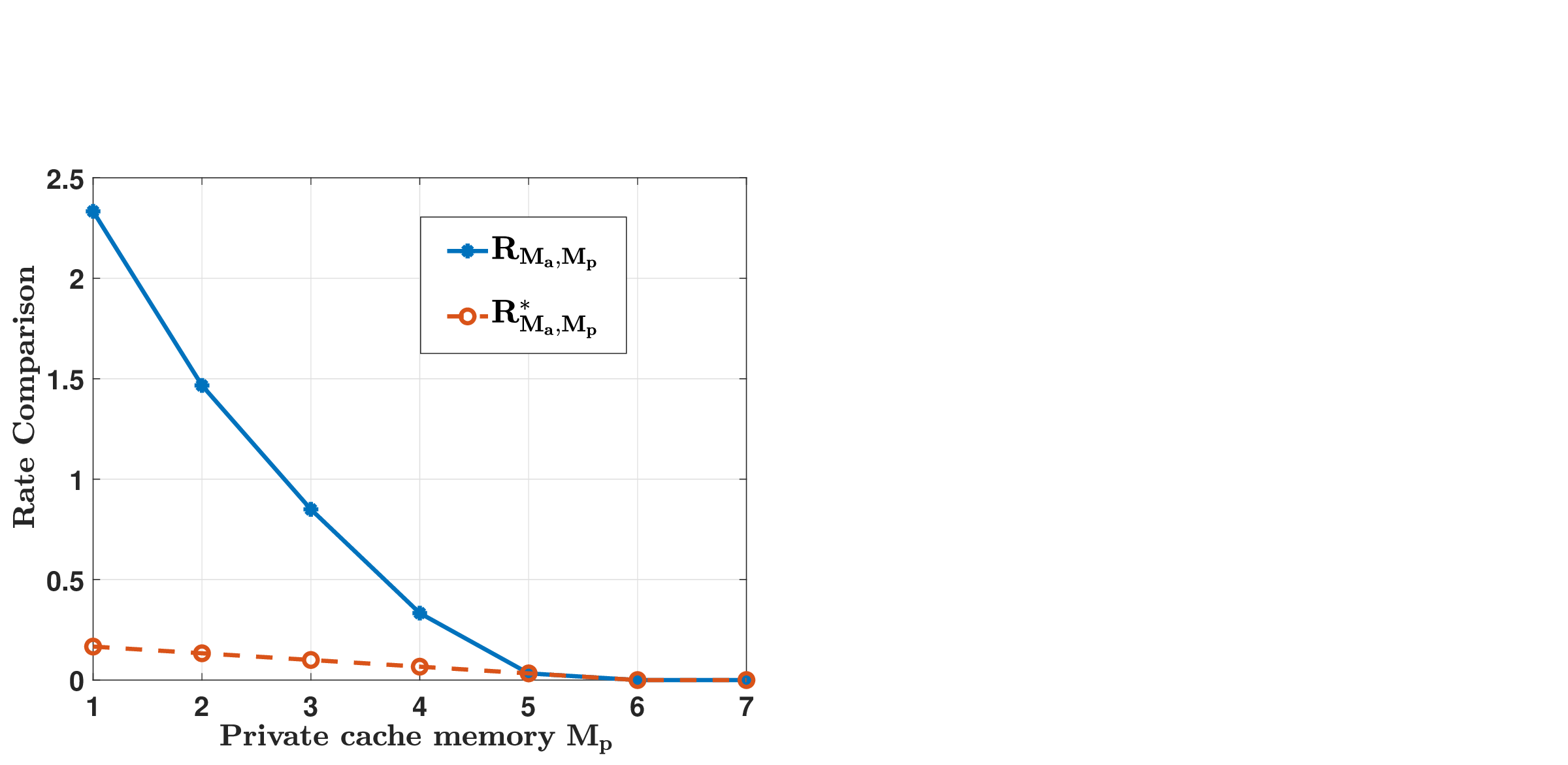}
	\caption{Rate vs $M_p$ comparison for the $(K=30,L=3,M_a=8,M_p,N=30)-$CW-MAP coded caching system.}
	\label{fig4}
	\includegraphics[width=\textwidth]{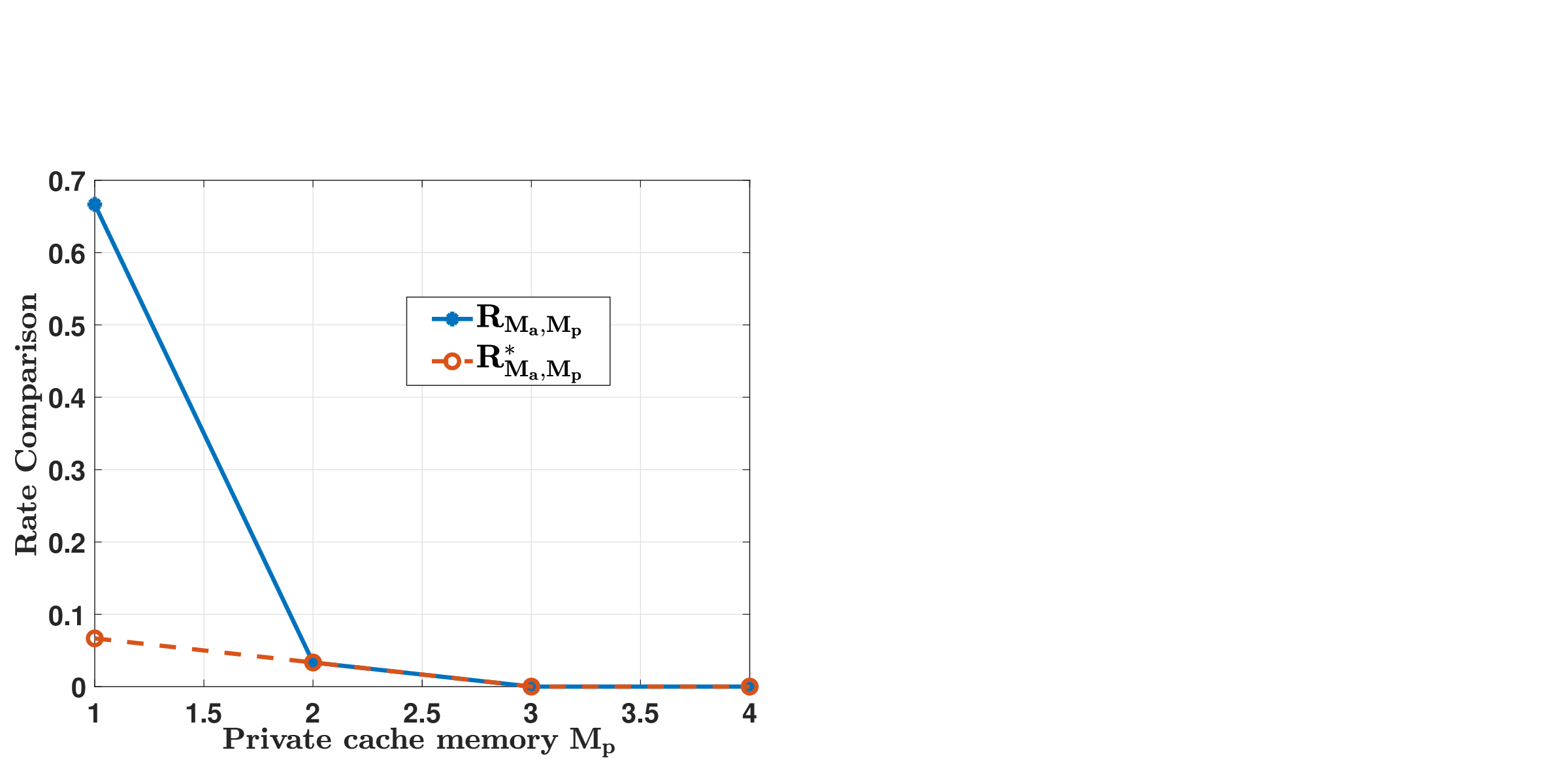}
	\caption{Rate vs $M_p$ comparison for the $(K=30,L=3,M_a=9,M_p,N=30)-$CW-MAP coded caching system.}
	\label{fig5}
\end{figure}

Fig. \ref{fig4}. characterizes the rate $R_{M_a,M_p}$ and the lower bound on $R^{\textasteriskcentered}_{M_a, M_p}$ for $L=3$ and $M_a=8$, for $M_p$ spanning the range of $[7]$. It can be observed that in Fig. \ref{fig4}., $R_{M_a,M_p}$ approaches optimality %$R^{\textasteriskcentered}_{M_a,M_p}$ 
as $M_P$ increases, achieving optimality for $M_p=5$.

Finally, Fig. \ref{fig5}. shows the variation of the rate-memory trade-off and the lower bound derived in Theorem \ref{thm1} for $L=3$ and $M_a=9$, with $M_p$ ranging from one to four. Fig. \ref{fig5}. demonstrates that $R_{M_a,M_p}$ moves closer to the lower bound %$R^{\textasteriskcentered}_{M_a,M_p}$ 
as $M_p$ increases, achieving optimality at $M_p=2$.

In essence, the optimality condition is achieved for $(M_a,L,M_p)=(6,3,11)$, $(M_a,L,M_p)=(7,3,8)$, $(M_a,L,M_p)=(8,3,5)$, and $(M_a,L,M_p)=(9,3,2)$, illustrating the concluding remark in Section \ref{ssec:optimality}. 
\section{Conclusions and Directions for Future Works}
\label{conclusions}
In this work, we introduced the CW-MAP coded caching setting where users connect to two types of caches, namely, private caches and access caches. For this setting, we proposed a centralized coded caching scheme, under uncoded placement, and characterized the rate-memory trade-off. A  
lower bound on the optimal rate, under any general placement, for the CW-MAP coded caching system using cut-set bound arguments\cite{CT} was provided. We proved the optimality of the proposed scheme for large memory regimes. Further, it was proven 
that optimality is achieved by several $(K, L, M_a, M_p, N)-$CW-MAP settings, all satisfying the same optimality condition and this was demonstrated using numerical comparison plots. There are a lot of interesting directions for future work in the CW-MAP setting some of which are as follows:

\begin{enumerate}
	\item Characterizing the rate of the achievability scheme for the memory regime $\gamma_p\geq \gamma_aL$.
	\item It can be observed from the numerical plot that the proposed scheme does not perform well in the low memory regime. Finding a coded caching scheme with better performance in the low memory regime will be an interesting problem.
	\item Improved lower bounds for the Cyclic Wrap-Around topology have been proposed in the literature\cite{NR}. It will be interesting to see if tighter lower bounds can be derived for the CW-MAP coded caching setting.
\end{enumerate}
\section*{Acknowledgment}
This work was supported partly by the Science and Engineering Research Board (SERB) of the Department of Science and Technology (DST), Government of India, through J.C Bose National Fellowship to Prof. B. Sundar Rajan.

\end{document}